\documentclass[times,tighten,twocolumn,xenglish]{aastex631}
\usepackage{lipsum}
\usepackage{physics}
\usepackage{multirow}
\usepackage{rotating}
\usepackage{graphicx}
\usepackage{subfigure}
\usepackage{color}
\usepackage{amsfonts}
\usepackage{mathrsfs}
\usepackage{babel}
\usepackage{amssymb}
\usepackage{wasysym}
\usepackage{verbatim}
\usepackage{multirow}
\usepackage{hyperref}
\usepackage{float}
\usepackage{ulem}
\usepackage{soul}
\usepackage{boondox-cal}
\def\cblue{\color{black}}
\usepackage{paralist}
\usepackage{appendix}
\usepackage{filecontents}

\begin{document} \title{Gleeok's Fire-breathing: Triple Flares of
AT~2021aeuk within Five Years from the Active Galaxy SDSS
J161259.83$+$421940.3}

\author[0000-0003-2024-1648]{Dong-Wei Bao$\dagger$}
\affiliation{National Astronomical Observatories, Chinese Academy of Sciences, 20A Datun Road, Chaoyang District, Beijing 100101, China, sgkmndks@gmail.com}%

\author[0000-0001-9457-0589]{Wei-Jian Guo$\dagger$}
\affiliation{National Astronomical Observatories, Chinese Academy of Sciences, 20A Datun Road, Chaoyang District, Beijing 100101, China, sgkmndks@gmail.com}%
\affiliation{Key Laboratory of Optical Astronomy, National Astronomical Observatories, Chinese Academy of Sciences, Beijing 100012, China}

\author[0000-0002-2419-6875]{Zhi-Xiang Zhang}
\affiliation{Department of Astronomy, Xiamen University, Xiamen, Fujian 361005, People's Republic of China, zhangzx@xmu.edu.cn}

\author[0000-0003-0202-0534]{Cheng Cheng}
\affiliation{Chinese Academy of Sciences South America Center for Astronomy, National Astronomical Observatories, CAS, Beijing 100101, China, chengcheng@nao.cas.cn}

\author[0009-0000-1228-2373]{Zhu-Heng Yao}
\affiliation{Key Laboratory for Particle Astrophysics, Institute of High Energy Physics, Chinese Academy of Sciences, 19B Yuquan Road, Beijing 100049, China}
\affiliation{School of Physical Science, University of Chinese Academy of Sciences, 19A Yuquan Road, Beijing 100049, People’s Republic of China}

\author[0000-0001-5841-9179]{Yan-Rong Li}
\affiliation{Key Laboratory for Particle Astrophysics, Institute of High Energy Physics, Chinese Academy of Sciences, 19B Yuquan Road, Beijing 100049, People's Republic of China}

\author[0000-0002-7330-4756]{Ye-Fei Yuan}
\affiliation{Department of Astronomy, University of Science and Technology of China, Hefei 230026, Anhui, China}

\author{Sui-Jian Xue}
\affiliation{National Astronomical Observatories, Chinese Academy of Sciences, 20A Datun Road, Chaoyang District, Beijing 100101, China, sgkmndks@gmail.com}%
\affiliation{Key Laboratory of Optical Astronomy, National Astronomical Observatories, Chinese Academy of Sciences, Beijing 100012, China}

\author[0000-0001-9449-9268]{Jian-Min Wang}
\affiliation{Key Laboratory for Particle Astrophysics, Institute of High Energy Physics, Chinese Academy of Sciences, 19B Yuquan Road, Beijing 100049, People's Republic of China}

\author[0000-0002-9390-9672]{Chao-Wei Tsai}
\affiliation{National Astronomical Observatories, Chinese Academy of Sciences, 20A Datun Road, Beijing 100101, China}
\affiliation{Institute for Frontiers in Astronomy and Astrophysics, Beijing Normal University,  Beijing 102206, China}
\affiliation{School of Astronomy and Space Science, University of Chinese Academy of Sciences, Beijing 100049, China}
\author[0000-0002-6684-3997]{Hu Zou}
\affiliation{National Astronomical Observatories, Chinese Academy of Sciences, 20A Datun Road, Chaoyang District, Beijing 100101, China, sgkmndks@gmail.com}%
\affiliation{Key Laboratory of Optical Astronomy, National Astronomical Observatories, Chinese Academy of Sciences, Beijing 100012, China}

\author[0000-0003-4280-7673]{Yong-Jie Chen}
\affiliation{Key Laboratory for Particle Astrophysics, Institute of High Energy Physics, Chinese Academy of Sciences, 19B Yuquan Road, Beijing 100049, People's Republic of China}

\author[0000-0002-0096-3523]{Wenxiong Li}
\affiliation{National Astronomical Observatories, Chinese Academy of Sciences, 20A Datun Road, Chaoyang District, Beijing 100101, China, sgkmndks@gmail.com}%
\affiliation{Key Laboratory of Optical Astronomy, National Astronomical Observatories, Chinese Academy of Sciences, Beijing 100012, China}

\author[0000-0003-4121-5684]{Shiyan Zhong}
\affiliation{South-Western Institute for Astronomy Research, Yunnan University, Kunming, 650500 Yunnan, People’s Republic of China}

\author{Zhi-Qiang Chen}
\affiliation{School of Physics and Technology, Nanjing Normal University, No. 1, Wenyuan Road, Nanjing, 210023, People's Republic of China}


%

\begin{abstract}

We present a noteworthy transient AT 2021aeuk exhibiting three
distinct optical flares between 2018 and 2023. It is hosted in a
{\cblue{radio-loud}} narrow-line Seyfert 1 (NLSy1) galaxy, with an
optical image showing a minor tidal morphology and a red
mid-infrared color ($\emph{W1}-\emph{W2}=1.1$). Flares II and III
exhibit rapid rises, and long-term decays ($\gtrsim 1000$ days)
{\cblue{with recurring after-peak bumps}}. The $\emph{g-r}$ color {\cblue{after subtracting the reference magnitude}} exhibited a rapid
drop and recovery during Flare II, followed by the minor after-peak
evolution in {\cblue{blue}} colors. We applied a canonical
tidal disruption event (TDE) fitting on the light curves which gives
a decay index $\emph{p}$ of $-2.99_{-0.14}^{+0.13}$ for Flare II and
$-1.61_{-0.65}^{+0.34}$ for Flare III. The blackbody fitting shows lower temperatures ($\sim 10^{3.8}$K) {\cblue{with minor after-peak evolution}}. {\cblue{The blackbody radius ($\gtrsim 10^{16}\ \rm cm$)}} and luminosity ($\sim 10^{45}\rm erg\ s^{-1}$) are larger than the typical TDE sample's. The time lag (in rest frame)
between ZTF $\emph{g-}$ and $\emph{r-}$band ($\rm
\tau_{g,r}=3.4^{+1.0}_{-0.9}$ days) significantly exceeds the
prediction from the standard accretion disk. Pre-burst spectra
reveal prominent Bowen fluorescence lines, indicating a vigorous or
potentially long-lasting process that enriches the local
metallicity. Additionally, we derived black hole masses of
$\log\it{M}_{\bullet}\rm=7.09^{+0.18}_{-0.31}\ \it{M}_{\odot}$ and
$\log\it{M}_{\bullet}\rm=7.52^{+0.08}_{-0.10}\ \it{M}_{\odot}$ using
$\rm H\beta$ and $\rm H\alpha$ emission lines. The variation and
recurring features of AT 2021aeuk {\cblue{are not likely induced by
the radio-beaming effect or Type-II superluminous supernova (SLSN-II), however, we cannot rule out the possibility of TDE or
enhanced active galactic nuclei (AGN) accretion process.}} The unusually high occurrence of
three flares within five years may also induced by the complex local
environment.

\end{abstract}


\section{Introduction}

The emergence of automated photometry surveys, such as the Catalina
Survey (CRTS; \citealt{Drake2009}), All-Sky Automated Survey for
SuperNovae (ASAS-SN; \citealt{Kochanek2017}), the Zwicky Transient
Facility (ZTF; \citealt{Masci2019}), the Asteroid Terrestrial-impact
Last Alert System (ATLAS; \citealt{Tonry2018}), and Pan-STARRS1
(PS1; \citealt{Chambers2016}), has significantly expanded the number
of optical transients in recent years. This allows for the further
study of a wide range of transient phenomena, such as blazars,
supernovae (SNe), TDEs, intrinsic
variability in AGN, and microlensing
events. The physical mechanisms underlying optical transients can be
complex and coupled together, which lead to intricate behaviors in
the multi-wavelength light curves and spectra, deviating from the
typical patterns for individual phenomena. By carefully analyzing
these transients and comparing them to known classes of objects, we
can gain valuable insights into the physical properties of their
progenitors and the surrounding environments that give rise to them.
Below, we present an overview of some optical variability patterns
and proposed theoretical models that can induce optical flares.

Blazars are distinguished with their relativistic jets pointed close
to our line of sight \citep{Wolfe1978}. This alignment leads to the
Doppler beaming, which significantly amplifies the emission across
observed wavelengths. The optical emission coming from the
synchrotron emission from relativistic electrons in the jet
\citep{Konigl1981, Urry1982} is polarized and usually displays
remarkable variability on timescales ranging from minutes to decades
(e.g., \citealt{Raiteri2003, Rani2010, Gopal2011, Goyal2018,
Chand2022}). Several models have been proposed to explain the
observed variability, including shocks within the jet
\citep{Marscher1985, Hughes1985}, turbulence and magnetic
reconnection (e.g., \citealt{Guofan2016, Kagan2015}), interaction
with external sources (e.g., \citealt{Araudo2010, Zacharias2017}),
and changes in viewing geometry (e.g., \citealt{Larionov2013}). More
comprehensive  reviews can be found in \citet{Bottcher2019,
Hovatta2019}.

TDEs (\citealt{Rees1988, Evans1989, Phinney1989}) happen when a star
wanders into the tidal radius of a black hole and is torn apart by
the gravitational force. Recent advancements in optical sky patrol
surveys have led to a significant increase in the number of TDEs
detected in the optical band \citep{vanvelzen2020, vanvelzen2011,
Gezari2012, van_ZTF_TDEsample, Hammerstein2023, Yao2023}. These
events exhibit a characteristic rapid rise to peak brightness
followed by a gradual decline. The observed slow decline is thought
to reflect the gradual fallback of debris streams. As these streams
fall back towards the black hole, they self-intersect (relativistic
precession effects are involved) and dissipate energy. This process
theoretically expects a power-law decline with a slope of $\sim
t^{-5/3}$ \citep{Rees1988, Phinney1989, Evans1989}. Notably,
observation of many TDEs has generally confirmed this theoretical
prediction. However, the measured temperature is consistently lower
than expected in the predictions of the standard disk theory, and
the post-peak minor evolution also deviates from what is expected.
This energy discrepancy poses a significant challenge to the
understanding of the physical mechanisms responsible for the optical
emission in TDEs. To address this controversy, researchers have
proposed several alternative models, including envelopes of
reprocessing layers \citep{Loeb1997, Guillochon2014, Roth2016,
Metzger2016, Metzger2022}, shock collision arising from stream
self-interaction \citep{shiokawa2015, Piran2015, Guohx2023}, outflow
involved to remove mass from the accretion disk \citep{Miller2015},
etc. These models offer potential explanations for the observed
discrepancy.

The TDE identification from automated optical surveys typically
excludes transients originating from galaxies known to host an AGN
to minimize confusion with AGN variability, which might lead to
overlooking TDEs within AGN environments. Recent models offer
different perspectives on the impact of AGN accretion disks on TDE
rates. Some research \citep{KarasV2007, MckernanB2022, Prasad2024,
WangYH2024, Kaur2024} suggests that the presence of an AGN accretion
disk can potentially lead to an enhanced TDE rate with secular
evolution of stellar orbits.

Recent studies have identified an increasing number of optical
transients that exhibit characteristics similar to TDEs but in a
rather long-term flaring state, which is likely associated with an
AGN hosted in the galaxy\citep{Kankare2017, Blanchard2017,
Trakhtenbrot2019, JiangN2019, Gromadzki2019, Neustadt2020,
Frederick2021, Cannizzaro2022, Makrygianni2023, Petrushevska2023,
WisemanP2024, HinkleJ2024}. These transients often exhibit a rapid
rise followed by a prolonged decay lasting over {\cblue{hundreds or
thousands of days}}, deviating from the typical light curve patterns
expected for canonical TDEs \citep{van_ZTF_TDEsample,
Hammerstein2023}. The physical origins of these transients remain
uncertain due to the similarities in their multi-wavelength
properties with other energetic phenomena such as {\cblue{TDE}},
SLSN-II, and intrinsic AGN
flares. CSS 100217, a transient event located within an
{\cblue{NLSy1}} galaxy with significant star formation activity,
initially drew attention due to its unusually bright optical flare,
reaching a peak magnitude of $\rm M_{\rm v(peak)}=-22.7\ mag$
surpassing the brightness of most known TDEs. Initially classified
as an extremely luminous type-II supernova (SN-II) based on its
early observations \citep{Drake2009}, CSS 100217 has recently been
reclassified as a potential TDE candidate \citep{Cannizzaro2022}.
This re-classification is primarily motivated by the fact that the
after-decay brightness of CSS 100217 is fainter than its pre-burst
brightness. This suggests that the central black hole in the host
galaxy is in a relatively low accretion state. A possible
explanation for this fainter after-decay brightness is the formation
of a cavity in the accretion disk during the TDE event of stars on
retrograde orbits \citep{MckernanB2022}. Similar fainter brightness
can be found in AT 2019brs \citep{Frederick2019}, AT 2021loi
\citep{Makrygianni2023} and AT 2019aalc \citep{veres2024} (Quick
comparison can be found in Fig.\ref{fig_longtermflares}). PS 16dtm,
residing in an NLSy1 galaxy of $M_{\rm BH}\sim10^6M_{\odot}$,
exhibited a double-peak plateau during the peak luminosity followed
by a smooth decline lasting over 7 years. Initially classified as an
SLSN-II \citep{Terreran2016, DongS2016}, PS 16dtm was later
reconsidered as a potential TDE candidate \citep{Blanchard2017,
Petrushevska2023}. Several other {\cblue{long-term-decay}}
transients, including AT 2019brs in \citet{Frederick2019} and
ZTF19aamrjar in \citet{WisemanP2024}, have also been identified at
pre-burst within galaxies known to host AGNs, {\cblue{which give
rise to the question of whether an AGN is involved in these
long-evolving behaviors.}} For some long-evolving transients in
AGNs, \citet{Trakhtenbrot2019} provided an alternative
interpretation that the flares are induced by intrinsic AGN
variability, potentially triggered by a sudden enhancement or
re-ignition of a supermassive black hole (SMBH). A key
characteristic of these ambiguous transients
\citep{Trakhtenbrot2019, Tadhunter2017, Blanchard2017,
Frederick2021, Makrygianni2023, veres2024} is the emergence of Bowen
fluorescence (BF) lines during the flaring phase \citep{Bowen1928,
Netzer1985} which is induced by the presence of intense UV
radiation. \citet{Dgany2023} suggest that the BF flares are rare
among SMBH-related transients.

Unfortunately, pre-burst observations are often unavailable for many
optical transients, making it difficult to definitively confirm the
presence of an AGN. PS1-10adi, for example, exhibits a slow,
long-term ($\sim$ 4 year) decay with a follow-up rebrightening phase
\citep{Kankare2017, JiangN2019}. The emergence of broad Balmer
emission lines, Fe {\sc ii} continuum, and Mg {\sc ii} features in
the spectrum of PS1-10adi can be attributed to TDE, AGN activity, or
an SN-II event. Some ambiguous/extremely nuclear transients
(ANTs/ENTs) have been identified in galaxies with tremendous energy
released but without any clear sign of AGN activity
\citep{Somalwar2023, HinkleJ2024, WisemanP2024}. It is worth noting
that some {\cblue{long-evolving transients}} even show recurring
flares \citep{JiangN2019, Payne2021, SoraisamM2022, Somalwar2023,
Makrygianni2023, veres2024}. The underlying physical processes
driving these new transients remain poorly understood, largely due
to the limited number of known examples. However, the discovery of
additional {\cblue{long-evolving transients}} will be crucial for
understanding the physical properties of progenitors and the
transient demography.

In this paper, we report a serendipitous discovery of an ambiguous
transient named AT 2021aeuk in SDSS J161259.83+421940.3, in which
three flares occurred within 5 years, and the second flare displayed
a long decline $\gtrsim$ 1000 days. We compile the archived light
curves and spectra aiming at placing constraints on the possible
physical origins of these flares. Section \ref{sec2} presents the
archived photometry data and calibration. Section \ref{sec3}
presents the measurement and analysis for light curves and spectra.
In Section \ref{sec4}, we discuss the possible physical mechanism
for AT 2021aeuk. In Section \ref{sec5}, we present the summary.
Throughout, we adopt a $\Lambda$CDM cosmology with $H_{0}= \text{67
km s}^{-1} \text{ Mpc}^{-1}$, $\Omega_{\Lambda}= 0.68$, and
$\Omega_{m}= 0.32$ (\citealt{Planck2020}).

\begin{figure}
\centering \includegraphics[width=0.45\textwidth]{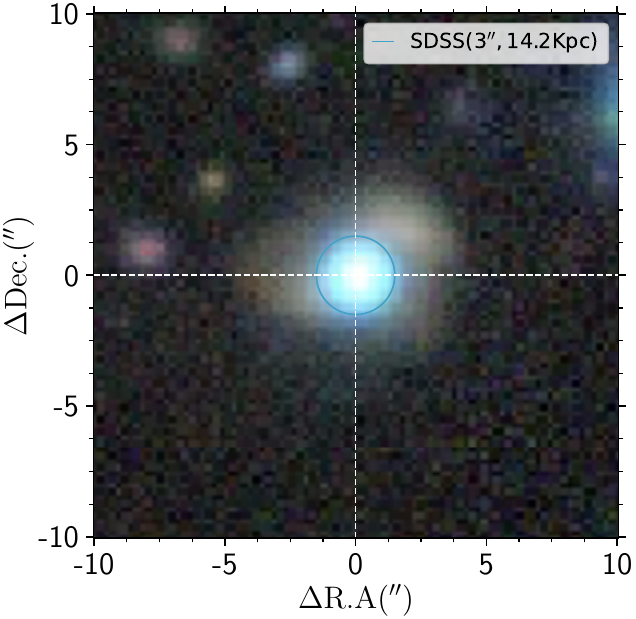}
\caption{The $g$-$r$-$i$ false-color image of AT 2021aeuk captured
by the Subaru/HSC \citep{2022PASJ...74..247A}.
The depth of the $r$ band image is about 26 AB mag. The SDSS fiber
aperture with diameter of 3$^{\prime\prime}$ is
denoted by the blue circle, corresponding to a scale of 14.2 kpc. }
\label{fig_subaru} \end{figure}

\begin{figure*}
\centering
\includegraphics[width=1\textwidth]{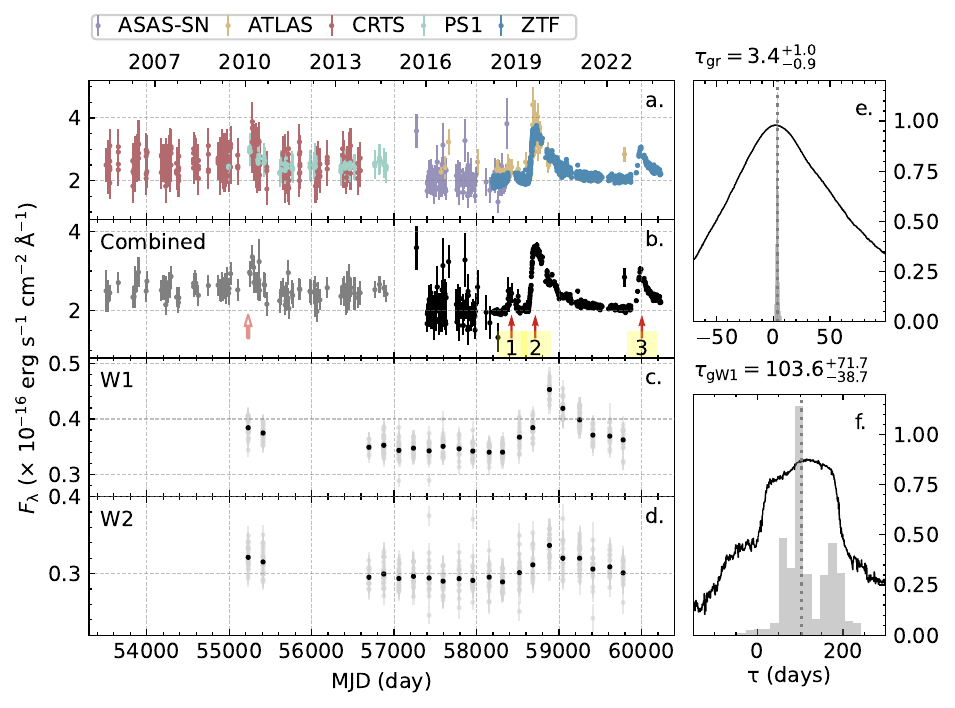}\\
\caption{The {\cblue{archived}} light curves in observer frame and time lag
measurements in rest frame for AT 2021aeuk. Panel a displays the
flux-scaled photometry data, compiled from ZTF, ASAS-SN, ATLAS,
CRTS, and PS1, each marked in different colors. Panel b
shows the combined light curve within 2 days. The red-filled arrows
with numbers mark three identified flares and the red unfilled arrow
around MJD 55200 marks a potential flare. The grey data points are
not flux-scaled with the black data points due to the sampling gap
around MJD 57000 (see Sec.\ref{sec_lightcurve}). Panels c and d
present $\emph{W1}$- and $\emph{W2}$-band photometric data from
AllWISE and NEOWISE surveys in grey points, along with the weighted
mean every six months marked in black. All fluxes are in the unit of
$10^{-16}~\rm erg\ s^{-1}\ cm^{-2}\ \AA^{-1}$. Panel e presents the
ICCF between ZTF $\it g$- and $\it
r$-band light curves and the centroid time lag $\tau_{\rm g,r}$ is
noted above Panel e. The grey histogram represents the distribution
of the centroid lags obtained from the Monte-Carlo calculation.
Panel f displays the CCF between the ZTF $\it g$-band and
$\emph{WISE}$ $\emph{W1}$-band. The analysis details can be found in
Sec.\ref{sec_timelag}.} \label{fig_lightcurve} \end{figure*}

\begin{figure*}
\centering
\includegraphics[width=1\textwidth]{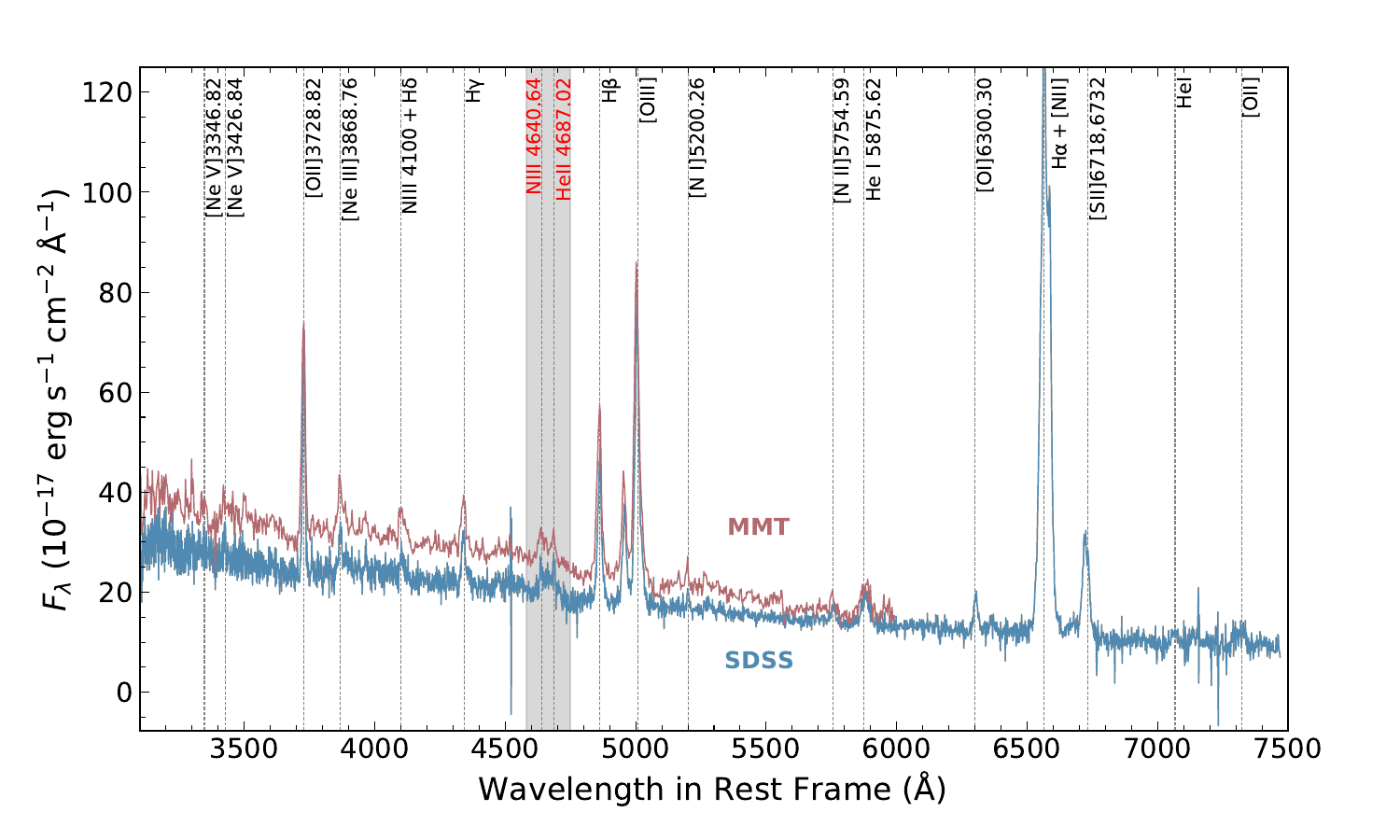}
\caption{The archival spectra of J1612 from MMT (in 1997) and SDSS
(in 2003). The dashed lines mark the prominent emission lines and
the grey span marks the BF lines (also see Sec.\ref{sec_AT}).}
\label{fig_spectrum} \end{figure*}

\section{Data Collection and Processing}
\label{sec2}

\subsection{Basic Information}

The host galaxy of AT 2021aeuk, SDSS J161259.83+421940.3 (hereafter
J1612), is classified as a radio-loud ($\log\ \it{R}^*\rm \sim 1.06$
from \citealt{Whalen2006}) NLSy1 galaxy\citep{White2000} at a
redshift of 0.234. It is optically blue while bright in the infrared
wavelengths with differential magnitudes $\emph{W1}$ - $\emph{W2}$ =
1.1, $W3$ - $W4$ = 2.569 archived from the $\emph{Wide-field Infrared
Survey Explorer}$ ($\emph{WISE}$) which suggests an origin of {\cblue{the reprocessing}}
from hot dust by AGN \citep{2005ApJ...631..163S}. The {\cblue{irregular morphology}}
revealed in the Subaru/Hyper Suprime-Cam (HSC) color image
(Fig.\ref{fig_subaru}) indicate that J1612 might have experienced a
galaxy merger event. The high-resolution Very Large Array (VLA)
observations at 5 GHz (beam size of $1''\times  0.42''$) show two
ambiguous radio cores with a separation of about 2 kpc
\citep{Berton2018}, which are not resolved by the HSC with a
resolution of $0.7''$.

\subsection{Light Curve and Calibration}\label{sec_lightcurve}

Various archived photometry data from different telescopes were
collected to extend the light curve, including ZTF, ASAS-SN, ATLAS,
CRTS, and PS1. The photometry data are presented in Fig.
\ref{fig_lightcurve}.

Due to different apertures and filters, the photometry data from
archived surveys require inter-calibration to present the long-term
variation. The inter-band flux scaling is conducted using the
Bayesian package
PyCALI\footnote{\url{https://github.com/LiyrAstroph/PyCALI}}
(\citealt{Li2014}), which fits the light curves with a damped random
walk process \citep{Kelly2009, Macleod2010, ZUying2013} and employs
Markov Chain Monte Carlo (MCMC) to recover the normalization
parameters and uncertainties for data from different telescopes. We
scaled all photometry data to ZTF $\it r$-band. However, we note
that there is no sampling overlap around MJD 57000 between the
datasets from CRTS/PS1 and ASAS-SN/ATLAS/ZTF. Therefore, the flux
scaling is not performed for datasets from CRTS/PS1 (see grey points
in Panel b in Fig.\ref{fig_lightcurve}). Nevertheless, this does not
affect our subsequent analysis since we focused on the light curves
after MJD 57000. ATLAS and ASAS-SN present highly consistent
variation with ZTF data, indicating the robustness of flux
calibration.

We also compiled the mid-infrared $\emph{WISE}$ data (see Fig.
\ref{fig_lightcurve}) and obtained 20 binned ({\cblue{within}} 50 days) {\cblue{data}} points
from 2010 to 2023 on $\emph{W1}$ (3.4$\rm \mu m$) and $\emph{W2}$
(4.6$\rm \mu m$) bands.

\subsection{Spectra of AT 2021aeuk} \label{sec_spectra}

The optical spectra (see Fig.\ref{fig_spectrum}) were obtained by
the Multiple Mirror Telescope (MMT) in 1997 (First Bright Quasar
Survey, \citealt{White2000}) and the Sloan Digital Sky Survey (SDSS)
in 2003. {\cblue{The spectra from MMT and SDSS are with the similar emission line features, but}} the minor difference in flux level which may be
caused by different fiber apertures or flux calibrations.
The detection of BF lines (see grey shaded part in
Fig.\ref{fig_spectrum}), particularly N {\sc iii}$\rm
\lambda\lambda$4640 \citep{Bowen1928, Netzer1985}, is uncommon in
typical AGN spectra \citep{vandenberk2001}. Their co-existence with
strong He {\sc ii} in both MMT and SDSS spectra implies the presence
of a powerful X-ray/UV photon source in AT 2021aeuk
\citep{Leloudas2019}.

\begin{figure*}
    \centering
    \includegraphics[width=1\textwidth]{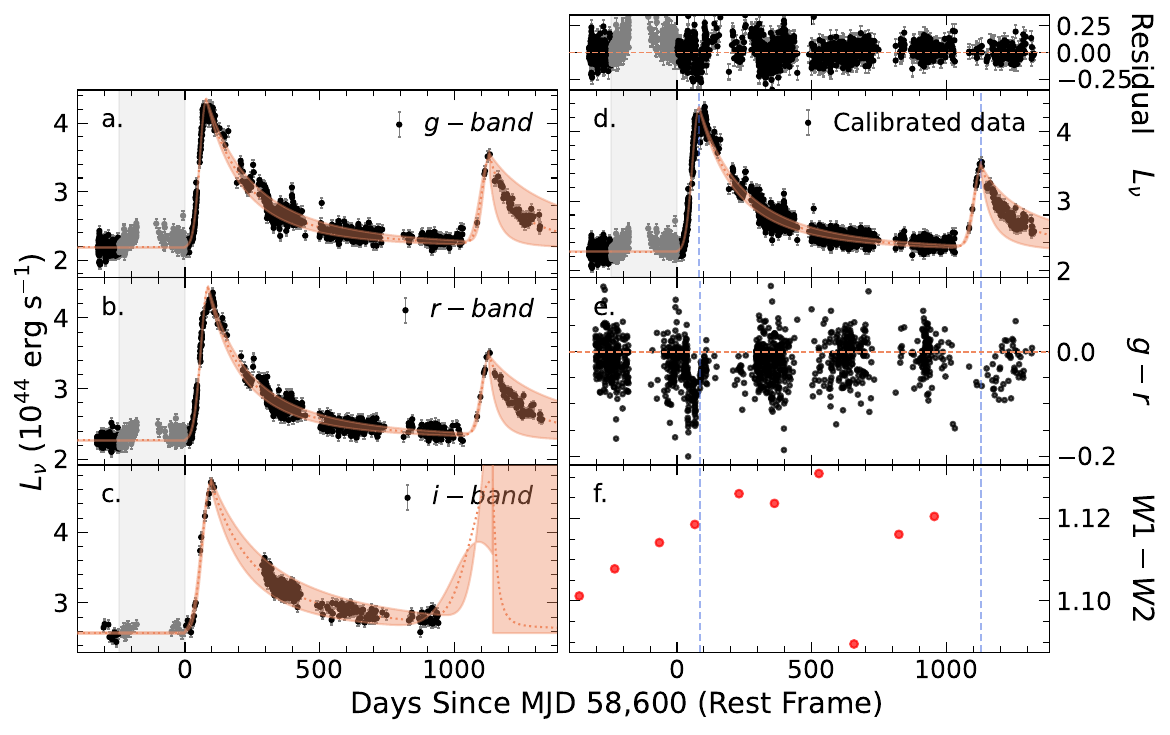}
    \caption{Panels a. to d. are the light curve fittings for ZTF $\emph{g-}$band, $\emph{r-}$band, $\emph{i-}$band, and combined light curves (scaled to $\emph{r-}$band) respectively. {\cblue{The x-axis is the days since MJD 55,600 in the rest frame and the y-axis is the luminosity with the galactic extinction corrected.}} The {\cblue{top right}} panel is the residual for combined light curve fitting. The dotted orange lines are the best-fit {\cblue{light curves}}, and the orange shaded regions are the 1$\sigma$ uncertainties. The vertical grey shaded parts are the masked region within 58300 $\textless$  MJD $\textless$ 58600. Panel e. is the $\emph{g-r}$ color evolution {\cblue{after subtracting the reference magnitude}}. Panel f. is the $\emph{W1-W2}$ color evolution. The blue vertical lines in Panels d., e., and f. {\cblue{locate at}} the peak luminosity of Flare II and Flare III in modeling. The x-axis is in the rest frame, and the galactic extinction corrections have been applied.}
    \label{fig_bestfit}
\end{figure*}

\section{Measurement and  Analysis}
\label{sec3}

\subsection{Light Curve Characterization}

We present optical and infrared light curves in
Fig.\ref{fig_lightcurve}. The optical light curves span
approximately 16 years, with the source remaining mostly quiescent,
exhibiting no significant flux fluctuation for the majority of the
time. While both CRTS and PS1
data suggest a possible rising feature around MJD 55200, limitations
in data quality make it difficult to determine its validity.
Therefore, we focus our analysis on the calibrated data after MJD
57000 for a more robust investigation of flares in AT 2021aeuk. In the past five years, AT 2021aeuk exhibited three optical
flares (hereafter Flares I, II, and III respectively as noted in
Fig.\ref{fig_lightcurve}) between Modified Julian Day (MJD) 58000
and 60400. Due to the discrepancy of data in the observational gap,
{\cblue{ the brightening feature around MJD 58,400 (Flare I) is ambiguous.}} Flare II {\cblue{exhibited}} rapid brightening
and a slow power-law decline lasting $\rm \sim 1000$ days (in rest
frame) after the peak, followed by Flare III displaying milder
brightening but a similar light curve {\cblue{pattern}}. The
$\emph{g-r}$ color evolution during the flaring is presented in
Fig.\ref{fig_bestfit}, in which {\cblue{the reference magnitudes were subtracted with $L_{v,0}$ obtained in Sec.\ref{sec_lcfit}}}. The color exhibited rapid drop and quick
recovery {{during}} Flare II followed by a minor evolution in {\cblue{blue}}
colors.

The $\emph{WISE}$ light curves reveal a broad, intense flare around
MJD 59000, coincident with Flare I and II seen in the optical light
curve, suggesting a wide distribution of hot dust. Due to the lack
of $\emph{WISE}$ data, the infrared echo to the Flare III around MJD
60000 remains unclear. The $\emph{W1-W2}$ color evolution during the
flaring is presented in Fig.\ref{fig_bestfit}, in which the
$\emph{W1-W2}$ maintains above 0.8 mag, consistent with the AGN
criterion in \citet{SternD2012}. The color generally shows minor
evolution considering the scattering of original data sets (see grey
points in Fig.\ref{fig_lightcurve}), and the drop around 530 days
may be induced by measurement uncertainties.

The physical mechanism behind these flares remains unclear. In this
section, we apply a generic quantitative analysis {\cblue{for}} TDEs to AT
2021aeuk to draw comparisons and explore the possible origins.

\subsubsection{Light Curve Fitting} \label{sec_lcfit} To quantify the rapid rise and slow decline in AT 2021aeuk's light
curves for Flares II and III, we adopt the fitting approach
described by \citet{van_ZTF_TDEsample}. This method utilizes a
combination of a {Gaussian} function for the rise phase and a power
law for the decay phase, as given by the following equation,
\begin{eqnarray} L_{\nu}(t)=L^{\rm{peak}}_{\nu} \times \begin{cases}
    e^{-(t-t_{\rm peak})^2/2\sigma^2}&\ t \leq t_{\rm peak}\\
    \left(\frac{t-t_{\rm peak}+t_{0}}{t_{0}}\right)^p&\ t>t_{\rm peak}
\end{cases} \label{eq_2} \end{eqnarray} where $L_{\nu}$ is the
luminosity, and $L^{\rm{peak}}_{\nu}$ is the peak luminosity at time
$t_{\rm peak}$.

The intrinsic variations of flares need to be isolated from the AGN
and/or host galaxy contributions during the quiescent periods.
However, the {\cblue{overlap of}} multiple flares on the light curve make this separation
uncertain. We added a constant luminosity component ($L_{\nu,0}$) in
our simultaneous fitting of Flares II and III (after masking out
Flare I). {\cblue{We performed fitting on individual ZTF bands and the combined light curve and the fitting is presented in Fig.\ref{fig_bestfit}.}} The fitting equations are provided below:
\begin{equation}
\label{eq_fit} L_{\nu}(t)= L_{\nu,0}+ \begin{cases}
L^{\prime}_{\nu,\rm {rise}}(t) &\    t \leq t^{\prime}_{\rm peak} \\
L^{\rm{\prime}}_{\nu,\rm {fall}}(t)+L^{\prime\prime}_{\nu,\rm {rise}}(t)&\ t^{\prime}_{\rm peak}< t \leq t^{\prime\prime}_{\rm peak} \\
L^{\prime}_{\nu,\rm {fall}}(t)+L^{\prime\prime}_{\nu,\rm
{fall}}(t)&\ t  > t^{\prime\prime}_{\rm peak} \end{cases}
\end{equation} where $t^{\prime}_{\rm peak}$  and
$t^{\prime\prime}_{\rm peak}$ are the time when Flare II and Flare III reached their peak luminosity respectively.

We employ the \texttt{emcee}\footnote{
\url{https://github.com/dfm/emcee}.} package
(\citealt{Foreman-Mackey2013}) to optimize the fitting. We adopt the
prior ranges presented in Table.\ref{tab_mcmcpriors} and the corner
plot is in Fig.\ref{fig_lccornerplot}. The optimal values are
determined using the median of the posterior distributions, and the
uncertainties are estimated using the $68.3\%$ confidence intervals.
We performed MCMC fits on both individual ZTF bands and the combined
light curve. The best-fitting parameters are summarized in Table
\ref{tab_fit}. The MCMC posterior distributions revealed a
significant degeneracy between $p$ and $t_{0}$. This degeneracy is
reflected in the wide range of best-fit values for $p$, which spans
from $-3.37$ to $-1.73$ when fitting individual bands. To ensure
better sampling robustness, we {\cblue{take}} the fitting parameters from
the combined calibrated light curve {\cblue{for further analysis}}. 
\begin{deluxetable}{cccc} \tablecolumns{4}
\vspace{+0.8cm} \tabletypesize{\footnotesize} \label{tab_mcmcpriors}
\tablehead{ \colhead{Parameters} & \colhead{Description} &
\colhead{Priors} & \colhead{Units} } \startdata
    $L_{\nu,0}$  & Luminosity in Quiescent time & [1,4] & $10^{44} \rm erg\ s^{-1}$ \\
    $\rm log L^{\rm peak}_{\nu}$ & Peak Luminosity & [0.01, 10] & $10^{44} \rm erg\ s^{-1}$ \\
    $t_{\rm peak1}$ & Peak Time of Flare II & [20, 200] & days \\
    $t_{\rm peak2}$ & Peak Time of Flare III & [1000, 1200] & days \\
    $p$ & Power-law Decay Index & [-10,0] & \\
    $\rm log \sigma$ & Gaussian Rise Time & [-3, 2] & days \\
    $\rm log t_{0}$ & Power-law Normalization & [-1, 5] & days
\enddata
\caption{Priors for MCMC Analysis. Except for $t_{\rm
peak}$, we adopted the same prior distributions for other parameters
of Flare II and Flare III.} \end{deluxetable}

In line with \citet{Yao2023}, $t_{1/2} = t_{1/2,\rm rise}+t_{1/2,\rm
decay}$ is introduced to characterize the light curve evolution {\cblue{timescale}}.
$t_{1/2,\rm rise}$ is the rising time from half-peak to peak
luminosity, while $t_{1/2,\rm decay}$ is the decaying time from peak
to half-peak luminosity. They are given in Table~\ref{tab_fit} as
well.

\begin{deluxetable*}{cccccccccc}
\tablecolumns{10}
\vspace{+0.8cm}
\tabletypesize{\footnotesize}
\tabcaption{\centering Best-Fitting Values for Light Curves.
\label{tab_fit}}
\colnumbers
\tablehead{
\colhead{Band} &
\colhead{Flare} &
\colhead{$L_{\nu,0}$} &
\colhead{$L^{\rm peak}_{\nu}$} &
\colhead{$t_{\rm peak}$} &
\colhead{$\sigma$} &
\colhead{$t_{0}$} &
\colhead{$p$} &
\colhead{$ t_{\rm 1/2,rise}$} &
\colhead{$ t_{\rm 1/2,decay}$}
 \\
 &   & ($10^{44} \rm erg\ s^{-1}$) &($10^{44} \rm erg\ s^{-1}$) & ($\rm days$)  & ($\rm days$) & ($\rm days$) &  & ($\rm days$)  & ($\rm days$)
}
\startdata
 & I & $2.27$   & \nodata &  \nodata & \nodata & \nodata & \nodata & \nodata & \nodata  \\
$g,r,i$ &  II & $2.27$  & $2.07^{+0.01}_{-0.01}$  & $82.86^{+0.38}_{-0.38}$  & $24.98^{+0.26}_{-0.25}$  &  $455.40^{+26.75}_{-25.02}$ & $-2.99^{+0.13}_{-0.14}$  & $30.45^{+0.49}_{-0.49}$ & $117.99^{+14.78}_{-13.30}$ \\
& III & $2.27$   &$1.19^{+0.03}_{-0.03}$ & $1129.01^{+2.02}_{-1.91}$ & $28.48^{+2.12}_{-1.98}$  &  $127.41^{+76.68}_{-38.73}$ & $-1.61^{+0.34}_{-0.65}$ & $33.97^{+3.95}_{-3.81}$ & $79.41^{+78.37}_{-50.93}$  \\
\hline
 &  I & $2.19$  & \nodata &  \nodata & \nodata & \nodata & \nodata & \nodata & \nodata  \\
 $g$ &  II  & $2.19$ & $2.16^{+0.01}_{-0.01}$  & $79.15^{+0.51}_{-0.49}$  & $23.93^{+0.34}_{-0.33}$  &  $519.22^{+47.14}_{-39.66}$ & $-3.37^{+0.20}_{-0.24}$  & $28.18^{+0.91}_{-0.91}$ & $118.67^{+23.23}_{-19.51}$  \\
&  III & $2.19$  &$1.29^{+0.05}_{-0.05}$ & $1129.32^{+3.48}_{-3.40}$ & $27.49^{+4.42}_{-3.88}$  &  $161.08^{+147.54}_{-71.70}$ & $-1.99^{+0.66}_{-1.30}$ & $34.41^{+7.52}_{-6.78}$ & $75.12^{+187.12}_{-62.95}$  \\
 \hline
&   I & $2.27$ & \nodata &  \nodata & \nodata & \nodata & \nodata & \nodata & \nodata   \\
$r$ &  II  & $2.27$ & $2.16^{+0.02}_{-0.02}$  & $85.76^{+0.67}_{-0.73}$  & $26.61^{+0.42}_{-0.45}$  &  $345.79^{+23.84}_{-22.21}$ & $ -2.49^{+0.12}_{-0.13}$  & $31.32^{+1.04}_{-1.05}$ & $110.79^{+16.51}_{-14.61}$    \\
 & III  &$2.27$  &$1.15^{+0.04}_{-0.04}$ & $1127.93^{+2.42}_{-2.42}$ & $26.08^{+2.36}_{-2.12}$  &  $146.25^{+151.60}_{-58.21}$ & $-1.73^{+0.49}_{-1.21}$ & $33.53^{+4.25}_{-4.03}$ & $86.15^{+203.74}_{-63.20}$  \\
 \hline
& I & $2.57$& \nodata &  \nodata & \nodata & \nodata & \nodata & \nodata & \nodata \\
$i$  &II  &$2.57$ & $2.14^{+0.04}_{-0.04}$  & $96.52^{+2.35}_{-2.25}$  & $32.42^{+1.47}_{-1.52}$  &  $385.57^{+73.24}_{-53.96}$ & $-2.30^{+0.25}_{-0.33}$  & $38.17^{+3.46}_{-3.47}$ & $135.78^{+56.26}_{-40.91}$  \\
& III & $2.57$& \nodata &  \nodata & \nodata & \nodata & \nodata & \nodata & \nodata
\enddata
\end{deluxetable*}
\vspace{-6mm}

\begin{figure*}
    \centering
    \includegraphics[width=0.8\textwidth]{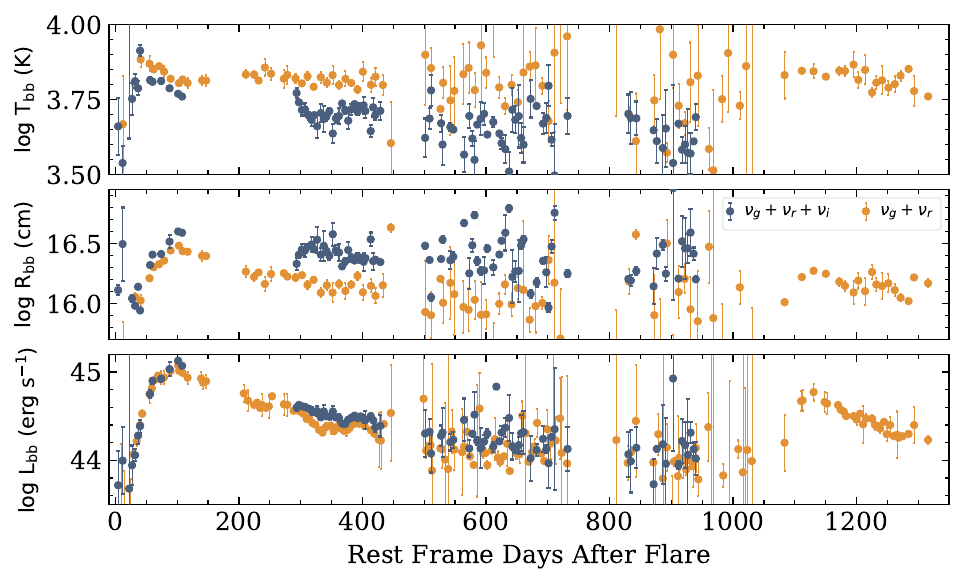}
    \caption{
    The blackbody temperature, radius, and luminosity for AT 2021aeuk in the top, middle, and bottom panels, respectively. The dark blue dots represent the results obtained from ZTF $g$-,$r$-,$i$-band data, while the orange dots correspond to the results derived from $g$-,$r$-band data. The $x$-axis is the same as Fig.~\ref{fig_bestfit}.
    }
    \label{fig_bbfit}
\end{figure*}

\begin{figure*}
    \centering
    \includegraphics[width=0.85\textwidth]{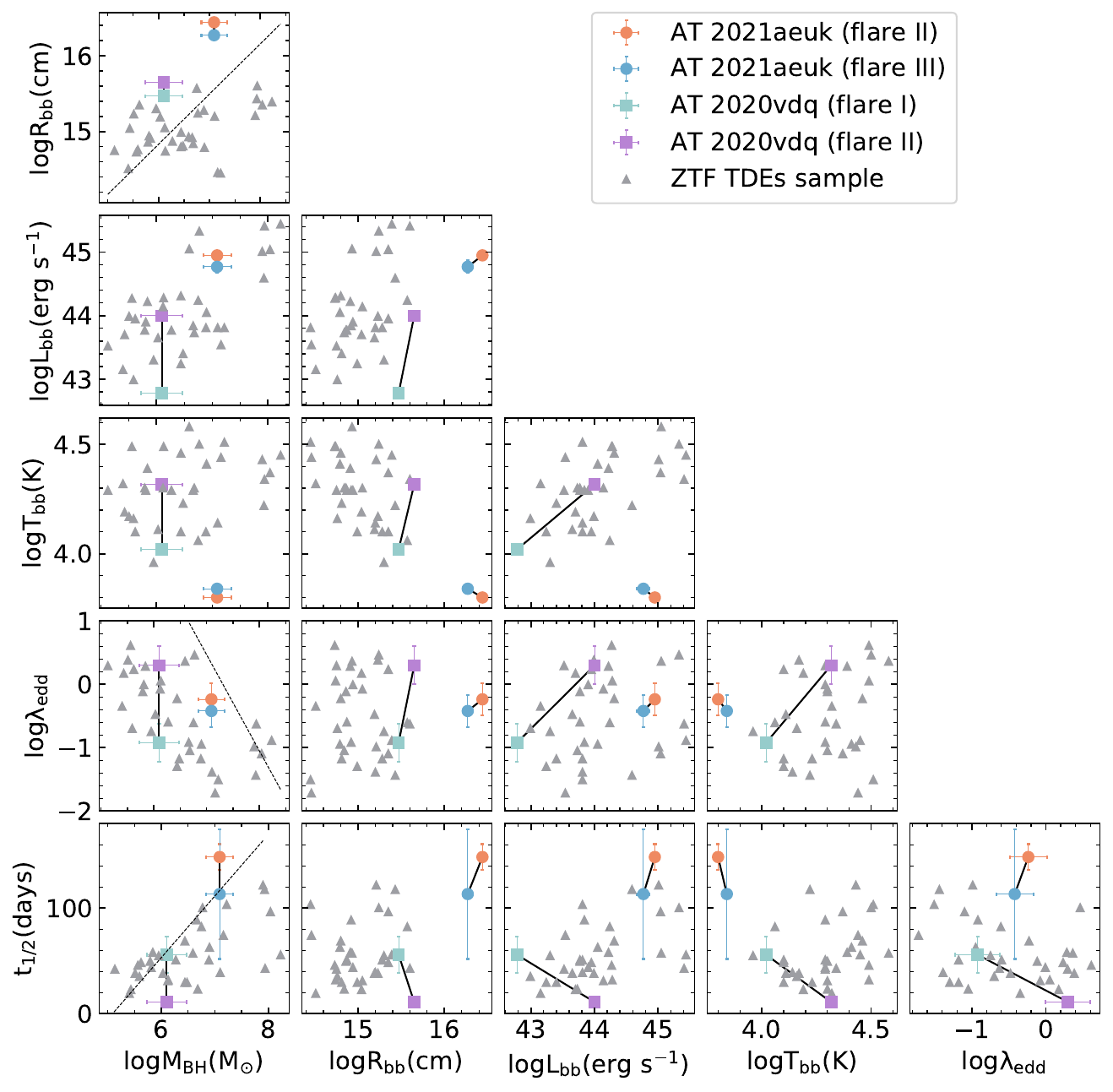}
    \caption{ The physical parameters of ZTF TDEs and two flares in AT 2021aeuk and AT 2020vdq. The flares of AT 2021aeuk are represented by orange and blue dots, while the flares of AT 2020 vdq are represented by green and purple dots. The TDE samples from  \cite{Yao2023} are presented in the grey dots. The grey dotted lines are the predicted scaling relationships $R_{\rm bb}\propto M_{\bullet}^{2/3}$ for the fiducial cooling envelop model from \cite{Metzger2016}, and $\lambda_{\rm Edd}\propto M_{\bullet}^{-3/2}$, and $ t_{1/2}\propto  M_{\bullet}^{1/2}$ (for the most tightly bound debris) from \cite{StoneN2013}, respectively.}
    \label{fig_TDEsample}
\end{figure*}

The UV/optical continuum of TDE is well described by thermal
blackbody radiation \citep{Gezari2021, Yao2023}. {\cblue{Hence,}} we applied
blackbody fitting to the intrinsic variations of flares in the ZTF
bands, {\cblue{to draw a comparison with the typical TDE sample.}} The parameter evolution is shown in Fig.~\ref{fig_bbfit} {{and the comparison with TDE sample is in Fig.\ref{fig_TDEsample}}}. The best-fit parameters at the peak luminosity are
summarized in Table~\ref{tab_fit2}.

\begin{deluxetable}{ccccc}
\tablecolumns{5}
\tabletypesize{\footnotesize}
\tabcaption{\centering The physical parameters inferred from the light curve.
\label{tab_fit2}}
\colnumbers
\tablehead{
\colhead{Flare} &
\colhead{$\rm  log\ L_{bb}^{peak} $} &
\colhead{$ \rm {log}\ T_{\rm bb}^{peak}$ } &
\colhead{$\lambda_{ \rm Edd}$} &
\colhead{$\rm log E_{ \rm bb}$}
 \\
  & ($\rm erg\ s^{-1}$) & (K) &  & ($\rm erg$)
}
\startdata
 I & \nodata & \nodata & \nodata &  \nodata \\
 II &  $44.95^{+0.03}_{-0.03}$ &  $3.80^{+0.01}_{-0.01}$ & $0.58$ &$52.30$ \\
 III & $44.77^{+0.10}_{-0.10}$ &$3.84^{+0.02}_{-0.02}$ & $0.38$ & $52.07$
\enddata
\end{deluxetable}
\vspace{-4mm}

\subsubsection{Time Lag Measurements}\label{sec_timelag}

The time delay between UV-optical-near-Infrared light curves is characteristic
evidence of AGN variability caused by reprocessing of radiation from
the inner accretion disk at different radii by the corona
\citep{Krolik1991, Cackett2007, Cackett2021}. In the standard
accretion disk model, the inter-band delay scales with wavelength as
$\rm \tau \propto \lambda^{4/3}$. For typical AGN variability, the
predicted time delay can be estimated based on the black hole mass
and luminosity, which is commonly $\sim$2-3 times smaller than the
observed time delay \citep{Fausnaugh2018, Kara2021, Guo2022}. As
such, UV-optical time lag measurement will help distinguish between
intrinsic AGN variability and some other mechanism, such as TDEs, as
the origins of flares in AT 2021aeuk.

The time spans (in rest frame) of ZTF $\it g$-band and $\it r$-band
are over 1500 days and the mean sampling cadences are $\sim$1.23
days and $\sim$0.94 days. We use the interpolated cross-correlation
function (ICCF; \citealt{Gaskell1986, Peterson1993}) to measure the
time delays in the rest frame (see Fig.\ref{fig_lightcurve}). The
centroid lags above 85\% of the coefficiency peak are adopted as
final lags, and the uncertainties are estimated from Flux
Randomization/Random Subset Sampling (FR/RSS) \citep{Peterson1998}.
We obtained a reliable time lag of $\tau_{\rm g,r}=3.4^{+1.0}_{-0.9}
\rm days$.

We also obtained an estimated time delay of $\tau_{g,\rm
W1}=103.6^{+71.7}_{-38.7}\ \rm days$ between the $g$-band and the
$\emph{WISE}$-$\emph{W1}$ light curve (Fig. \ref{fig_lightcurve},
panel f). The mean sampling cadence is $\sim 146.85\pm13.39$ days,
indicating the sample deficiency which may introduce uncertainties in
the measurement of $\emph{g-W1}$ time lag.

\subsection{Spectral Fitting and BH Properties} \label{sec_fittingandBH}

We employ
DASpec\footnote{\url{https://github.com/PuDu-Astro/DASpec}} to
perform spectral decomposition on the SDSS spectrum and estimate
black hole mass and accretion rate. The continuum is composed of a
power-law AGN continuum, galaxy template, and Fe {\sc ii}
pseudo-continuum. Emission lines were modeled with a single or
double Gaussian depending on the line profiles. Double Gaussians
were used for broader lines [O {\sc iii}]$\rm \lambda\lambda\
4959,5007$ potentially affected by outflows. The full-width at
half maximum (FWHM) and line shift parameter {\cblue{are tied}} together to account for
the common broadening mechanism. The flux ratio between $\rm H\beta$
and [O {\sc iii}]$\rm \lambda\ 5007$ was fixed at the canonical
value of 0.1 for NLSy1 fitting \citep{Veilleux1987, Vaona2012,
HuangY2019}. Similarly, the FWHM and line shift of the [N {\sc
ii}]$\rm \lambda\lambda\ 6548,6583$ and [S {\sc ii}]$\rm
\lambda\lambda\ 6716,6730$ doublets were tied together. {\cblue{Additional Gaussian components were}} used to {\cblue{describe}} the broad $\rm
H\beta$ and $\rm H\alpha$ emission arising from the broad line
region (BLR).

Due to weak galaxy absorption features and Fe II pseudo-continuum,
their contribution to the 5100 Å luminosity is negligible. We
obtained a 5100 Å luminosity of log$L_{5100} = 44.29\pm 0.01\ \rm
erg\ s^{-1}$, and we estimated a BLR size of {\cblue{$\sim
47.80_{-5.06}^{+5.22}\ \rm It\ days$}} based on the R-L relationship
from \citealt{Bentz2013}. We measured the FWHM of the broad $\rm
H\beta$ ($\rm 1091\pm 269\ km\ s^{-1}$) and $\rm H\alpha$ ($\rm
1788\pm 149\ km\ s^{-1}$) emission lines. Under the virial
assumption, we obtained the BH mass \citep{Blandford1982,
Peterson1993}, \begin{equation} M_{\bullet}  =  f_{\rm
BLR}\frac{R_{\rm BLR} V^2}{G} \end{equation} where $R_{\rm BLR} =  c
\tau_{\rm BLR}$ is the responsivity-weighted radius of the BLR,
$\tau_{\rm BLR}$ is the time lag, $c$ is the speed of light, $V$ is {\cblue{the}}
 FWHM of the emission lines from the spectra, $G$ is the
gravitational constant, and $f_{\rm BLR}$ is a scaling factor taken
as 1.12 for NLSy1 \citep{Woo2010}. The BH masses estimated from $\rm
H\beta$ and $\rm H\alpha$ are
$\log(M_{\bullet})=7.09_{-0.31}^{+0.18} M_{\odot}$ and
$\log(M_{\bullet})=7.52_{-0.10}^{+0.08}\ M_{\odot}$, respectively.
These masses generally agree with previous estimates for this
source: $M_{\bullet}=10^{7.3}\ M_{\odot}$ based on $\rm H\alpha$
\citep{GreeneJ2007} and $M_{\bullet}=10^{6.8} M_{\odot}$ based on
$\rm H\beta$ \citep{CraccoV2016}. We also gave the estimation of
dimensionless accretion rate \citep{FrankJ2002, NetzerH2013,
WangJM2014}, \begin{equation}
    \dot{\mathscr{M}} = 20.1\frac{\mathcal{l}_{44}}{\cos\ i}m_7^{-3/2}
\end{equation} where $\mathcal{l}_{44}=L_{5100}/10^{44}\rm erg\
s^{-1}$, and $m_7=M_{\bullet}/10^7 M_{\odot}$. $i$ is the
inclination of the disk to the line of sight, and we take $\rm cos\
i=0.75$ \citep{DuPu2016}. We {\cblue{obtain}} the accretion rate of
$\sim$1.74 and $\sim$0.88 from BH masses derived from $\rm H\beta$
and $\rm H\alpha$, respectively. To assess the accretion power for
the flares, we eliminate the intrinsic AGN and/or host galaxy
contribution $L_{\nu,0}$ and calculated the Eddington ratio
($\lambda_{\rm{Edd}} = L_{\rm bb}^{\rm peak}/L_{\rm Edd}$) of
$\sim$0.58 and $\sim$0.38 at the peak luminosity of Flare II and
III, where $L_{\rm Edd}\equiv(M_{\rm BH}/M_{\rm
\odot})\times1.25\times10^{38}\rm erg\ s^{-1}$ \citep{Yao2023}. The
results are presented in Table~\ref{tab_fit2}.

\section{DISCUSSION}
\label{sec4}

\subsection{Scenario of Radio-Beaming Effect}

The estimated occurrence rate of radio-loud AGNs among {\cblue{NLSy1}} is $\sim
4 \%$ \citep{Komossa2006, zhouhongyan2006, Rakshit2017, Singh2018}.
J1612, as an {\cblue{NLSy1}}, is classified as a flat-spectrum radio quasar
(FSRQ) with a loudness parameter of $\rm log\ R^* \sim 1.06$. The
radio-loud nature and the VLA resolved features \citep{Berton2018}
of J1612 suggest the potential presence of relativistic jets. The
jets are a hallmark of blazars, where the emission is
Doppler-boosted due to the beaming effect along the jet direction.
In J1612, the optical emission might be similarly influenced by this
beaming effect. We explore the possibility of the radio-beaming
effect contributing to the optical flares in AT 2021aeuk. However,
based on the following evidence, we argue that jet-induced emission
likely plays a minor role in the observed optical flares,
\begin{itemize} \item \textbf{Light Curve Morphology.} Blazars are
renowned for their highly variable optical emission, exhibiting
rapid rise and fall times with a symmetrical light curve shape
\citep{Raiteri2003, Rani2010, Gopal2011, Goyal2018}. This behavior
is distinct from the observed optical flares in AT 2021aeuk, which
displayed a fast rise but a slow, monotonical decline. \item
\textbf{Optical-Infrared Time Delay.} Blazars generally exhibit minimal
time lags between their optical and near-Infrared emission
variations, typically on the order of days or less
\citep{Paltani1997, Neugebauer1999, Raiteri2006, Bonning2012,
Liaonenghui2019, Lyujianwei2019, Perger2023}. In contrast, the
observed delay of about 120 days between the optical and
$\emph{g-W1}$ (3.4 $\mu$m) flares in AT 2021aeuk (see
Fig.\ref{fig_lightcurve}) is significantly longer than what can be
explained by standard blazar infrared emission processes, which can
instead be explained by the dust reprocessing from the dusty torus.
\item \textbf{Long-Term Quiescent State.} The blazar state of FSRQs
is considered to last for decades, with occasional state transitions
on year-long timescales \citep{Chand2022}. However, the 16-year
light curve suggests that AT 2021aeuk spends most of this period in
a low state, exhibiting major flares only during recent years.
Besides, we corrected the extinction of the SDSS photometry
($R_{V}=3.1$, $E(B-V)=0.014$, \citealt{Fitzpatrick1999PASP}) and
yielded magnitudes of $r$ = 17.8 and $g$ = 18.2, which are
consistent with the ZTF photometry obtained during the quiescent
state. \item \textbf{Absence of $\gamma$-ray counterpart.} Blazars
typically exhibit a strong correlation between optical and
$\gamma$-ray emission variations, with $\gamma$-rays often leading
the optical changes by a small time delay (e.g.,
\citealt{raniB2013}). Interestingly, no $\gamma$-ray counterpart was
detected by the $\it Fermi$ Large Area Telescope (LAT) during the
flaring periods in AT 2021aeuk (private communication with Jun-Rong
Liu). Although ``orphan flares" have been observed in blazars, they
only appear in one band ($\gamma$-ray band or optical band) and do
not have counterparts in the other band. The mechanism can be the
accretion-disk outbursts or stochastic fluctuation of multiple
zones, but the {\cblue{orphan flares}} usually come with rather lower amplitudes
\citep{Liodakis2019}. \end{itemize}

\subsection{Scenario of TDEs} \label{Sec_senarioofTDEs} Large-scale
optical sky surveys have significantly boosted the number of
discovered UV-optical TDEs. However, TDE identification remains
challenging due to the uncertainties surrounding their physical
origin and the relatively limited number of confirmed cases.
Consequently, the current criteria for identifying TDEs are
primarily based on observable features, and we investigate the
criteria proposed by \citet{vanvelzen2019, vanvelzen2020} below. It
should be noted that some TDEs exhibiting atypical characteristics {\cblue{which}}
might not satisfy the established criteria. \begin{itemize}
    \item \textbf{Nuclear.} The observed infrared echo (see Sec.\ref{sec_timelag} and Fig.\ref{fig_lightcurve}) {\cblue{indicates the thermal reprocessing of dusty torus ignited either by AGN accretion \citep{Barvainis1987} or nuclear transients, like supernova or TDE \citep{Luwb2016, DouLM2017, Jiang2017, Yang2019}}}. {\cblue{However,}} the observed time lag ($\sim103.6_{-38.7}^{+71.7}$ days), while uncertain, is generally {\cblue{smaller than the expected light traveling time in the mid-infrared band ($\sim 325.0\ \rm days$) induced by the typical AGN radiation with log$L_{5100} = 44.29\pm 0.01\ \rm erg\ s^{-1}$ \citep{Koshida2014, minezaki2019, Lyujianwei2019,chen2023}. This suggests that the distance between the UV/optical emitter and the dusty region is relatively smaller than the expected distance between the dusty torus and the accretion disk. The infrared echo is more likely induced by nuclear transients like TDE or supernova.}} 
    \item \textbf{Spectroscopy} The spectra from MMT (1997) and SDSS (2003) (see Fig.\ref{fig_spectrum}) are not taken during the flare. However, they both exhibited prominent N {\sc iii} $\lambda$4640, which suggests the unusual local metallicity abundance. A TDE with a high-mass star can be attributed to the abundance (e.g., \citealt{kochanek2016, liuxin2018}), and the line width of N {\sc iii} $\lambda$4640 and He {\sc ii} $\lambda$4686 may be highly broadened through electron scattering during the flares \citep{RothN2018, Leloudas2019}. Since we do not have spectra during the flares, we cannot distinguish them from other physical mechanisms (as stated in Sec.\ref{sec_AT}).
    \item \textbf{Light Curve Morphology}. Similar to typical TDE light curves, both Flare II and Flare III in AT 2021aeuk exhibit a rapid rise followed by a slower decline. The $t_{\rm 1/2, rise}$ for Flare II ($\rm 30.45_{-0.49}^{+0.49}$ days) and Flare III ($\rm 33.97_{-3.81}^{+3.95}$ days) are generally consistent and the $t_{\rm 1/2, decay}$ for Flare II ($\rm 117.99_{-13.30}^{+14.78}$ days) is longer than that for Flare III ($\rm 79.41_{-50.93}^{+78.37}$ days). However, during the {\cblue{decay}} phase, the power-law index for Flare II ($p=-2.99_{-0.14}^{+0.13}$) is steeper than the theoretical value of $-5/3$. On the other hand, Flare III's index ($p=-1.61_{-0.65}^{+0.34}$) falls closer to the expected value {\cblue{which may vary with new coming data afterward. Besides, Flare II and III exhibited post-peak bumps during the decay phase, which may be explained as the late-time forming of the disk during the TDE \citep{Guohx2023}}}.
    \item \textbf{Color and Temperature}. The $g-r$ color (see Fig.\ref{fig_bestfit}) during the flare remains {\cblue{blue}}, and it exhibited a rapid drop and quick recovery {\cblue{during}} Flare II followed by minor evolution later, {\cblue{which are consistent with}} the behaviors of typical TDEs \citep{van_ZTF_TDEsample, Yao2023}. The blackbody temperature (see Fig.\ref{fig_bbfit}) from ZTF $g-$, $r-$band fitting exhibited no significant cooling trend, which is consistent with the behaviors in most UV/optically-selected TDEs \citep{van_ZTF_TDEsample, Gezari2021, Yao2023}. The peak temperatures measured for Flare II ($\rm logT_{bb}^{peak}\sim3.80_{-0.01}^{+0.01}$) and Flare III ($\rm logT_{bb}^{peak}\sim3.84_{-0.02}^{+0.02}$) are significantly lower than the values typically observed in UV/optical TDEs ($\rm >1.2\times10^4K$, see Fig.\ref{fig_TDEsample}, \citealt{vanvelzen2020, Yao2023}). However, the lack of UV photometry and internal extinction {\cblue{correction}} may introduce some uncertainties in blackbody fitting. These behaviors also disfavor the SN origin which will show post-peak color reddening and prominent temperature cooling. It should be noted that the minor temperature post-peak drop may be induced by the peak variation inconsistency between ZTF $g-$, $r-$ bands (see Panels a and b in Fig.\ref{fig_bestfit}).
\end{itemize} The current analysis is insufficient for a definitive
TDE classification based on the criteria discussed above. To gain
further insights {\cblue{of the origin}}, we compared their disruption properties  (see Fig.\ref{fig_TDEsample})
with those of well-established TDEs \citep{Yao2023}. The blackbody
radius of AT 2021aeuk {\cblue{($\gtrsim 10^{16}\rm\ cm$)}} is larger than it typically is in other
TDEs. {\cblue{It also}} deviates from the
predicted $R_{\rm bb}\propto M_{\bullet}^{2/3}$ for the fiducial
cooling envelope model {\cblue{(see Fig.\ref{fig_TDEsample}, \citealt{Metzger2016})}}. The half-max during
($\rm t_{1/2}$) is much longer than the typical TDE samples.
Interestingly, the event {\cblue{appears}} to follow the theoretical
predictions \citep{StoneN2013} for $\lambda_{\rm Edd}\propto
M_{\bullet}^{-3/2}$ and $t_{1/2}\propto M_{\bullet}^{1/2}$.

\subsection{Scenario of AGN Variation}
\label{sec_AGN_typical_va}

\begin{itemize}
    \item \textbf{Variability Pattern.} The typical AGN variability displays stochastically $\sim 10\%$ {\cblue{ with symmetric flares }} and can be effectively produced by the damped random walk (DRW) model \citep{Kelly2009, Kozlowski2010, Macleod2010, ZUying2013}. {\cblue{Besides,}} the typical variation of NLSy1 is systematically smaller \citep{AiYL2010}. Hence, the asymmetric flares of AT 2021aeuk, characterized by rapid smooth rise and slow decay, are unlikely well-described by the {\cblue{typical AGN variation}}.
    \item \textbf{Inter-band Time Lag.} The time lag of typical AGN variation between $g-$, $r-$ bands (see Sec.\ref{sec_timelag}) based on the standard disk model is $\tau^{\rm ss}_{\rm g,r}=0.28\ \rm\ days$ which is $\sim 12.1$ times shorter than the observed one ($\rm \tau_{g,r}=3.4_{-0.9}^{+1.0}$ in days). The substantially long delay is reported in some TDEs detected by ZTF \citep{Guo2023}. {\cblue{Besides, the time lag between $\emph{g-}$ and $\emph{W1-}$ disfavors the typical AGN-induced torus echo (see Sec.\ref{Sec_senarioofTDEs}).}}
\end{itemize}

The asymmetric variation, the long {\cblue{$\emph{g-r}$}} time lag, and the {\cblue{short $\emph{g-W1}$}} time lag, and the
distinct inter-band variation patterns all suggest that the physical
mechanisms behind the optical flares in AT 2021aeuk are insufficient
to be explained by a typical AGN radiation process.

The changing-state phenomenon in AGNs, where the broad emission
lines emerge or go, can exhibit dramatic changes in optical
brightness (typically months to years, \citealt{Ricci2023}). This is
commonly explained by the change in accretion rate caused by disk
instabilities \citep{LinDNC1986, Balbus1991} or major disk
perturbation (e.g., TDE, \citealt{Merloni2015}). Changing-state
behaviors seem to prefer to reside in AGNs with low Eddington ratio
with $\rm logL/L_{Edd} \lesssim -1.0$ \citep{Macleod2019,
GreenPJ2022, GuoWJ2024}, which is smaller than the Eddington ratio
{\cblue{of}} J1612 ($\sim 0.88$, see Sec.\ref{sec_fittingandBH}). However, we
do not have {\cblue{on-burst}} spectra and cannot verify whether the broad emission lines have gone through changing-look behaviors.

\subsection{Scenario of Ambiguous Transients}
\label{sec_AT}

\begin{figure*}
    \centering
    \includegraphics[width=1\textwidth]{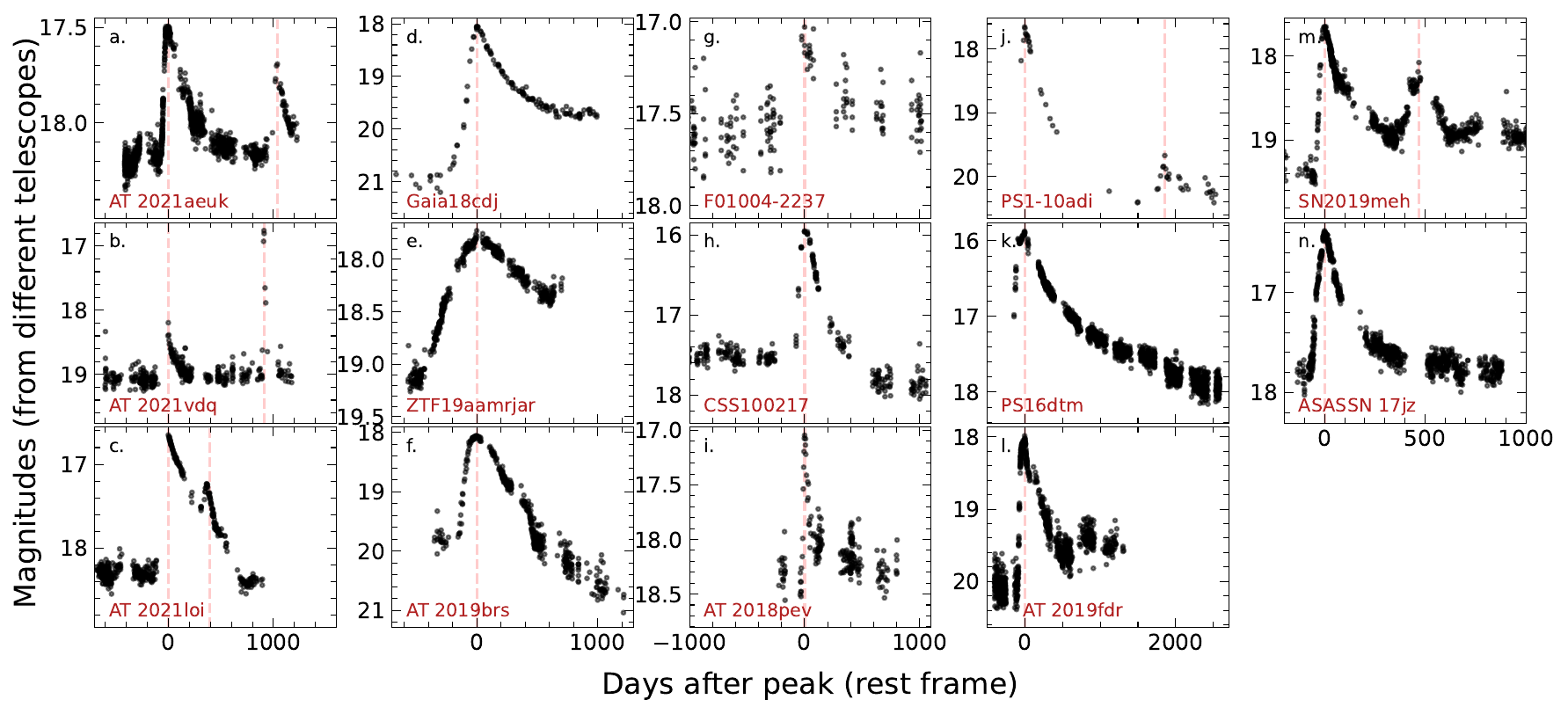}
    \caption{Long-term flares in the rest-frame day since flare. Panel a. is ZTF $g$-band data for AT 2021aeuk in this paper. Panel b. is ZTF $g$-band data for AT 2021vdq (pTDE candidate in \citealt{Yao2023, Somalwar2023}); Panel c. is ZTF $g$-band data for AT 2021loi (BFF candidate in \citealt{Makrygianni2023}); Panel d. is data for Gaia18cdj from Gaia Alert Transient Stream (ENT, TDE candidate in \citealt{HinkleJ2024}); Panel e. is ZTF $g$-band data for ZTF19aamrjar (ANT in \citealt{WisemanP2024}); Panel f. is  ZTF $g$-band data for AT 2019brs (BFF candidate in \citealt{Frederick2021}); Panel g. is $\emph{V}$-band data from CRTS for F01004-2237 (TDE candidate in \citealt{Tadhunter2017}, BFF candidate in \citealt{Trakhtenbrot2019}); Panel h. is $\emph{V}$-band data from CRTS for CSS100217 (SN II candidate in \citealt{DrakeAJ2011}, TDE candidate in \citealt{Cannizzaro2022}); Panel i. is  ZTF $g$-band data for AT 2018pev (BFF candidate in \citealt{Frederick2021}); Panel j. is $\emph{V-}$band data for PS1-10adi (SLSN-II or TDE candidate in \citealt{Kankare2017, JiangN2019}); Panel k. is Pan-STARRS data for PS16dtm (TDE candidate in \citealt{Blanchard2017}); Panel l. is ZTF $g$-band data for AT 2019fdr (TDE candidate in \citealt{Frederick2021}); Panel m. is ZTF $g$-band data for SN 2019meh (SLSN-II candidate in \citealt{NichollM2019, SoraisamM2022}); Panel n. is ATLAS data for ASASSN-17jz (SN IIn candidate in \citealt{Holoien2022ApJ}.)}
    \label{fig_longtermflares}
\end{figure*}

The light curve evolution and pre-burst spectral properties of AT
2021aeuk exhibit notable deviations from the typical characteristics
observed in most confirmed canonical transients {\cblue{induced by the single physical mechanism}}. We also compare AT 2021aeuk to several
{\cblue{ambiguous}} transients {\cblue{(see Fig.\ref{fig_longtermflares})}} , including Bowen
fluorescence flares (BFFs, \citealt{Trakhtenbrot2019}), ANTs/ENTs
\citep{HinkleJ2024, WisemanP2024}, and SLSNe-II \citep{Miller2009,
GezariS2009, Inserra2018}. It is important to
note that definitive classification remains challenging due to the
limited number and the incomplete demography of
these phenomena. \begin{itemize}
    \item \textbf{Light Curve Properties}. Flare II in AT 2021aeuk exhibits a prolonged decay phase (lasting over 1000 days). The blackbody temperature associated with Flare II/III is lower than typically seen in TDEs, additionally, the peak luminosity and blackbody radius are higher (see Fig.\ref{fig_TDEsample}). AT 2021loi, {\cblue{classified as}} a BFF \citep{Makrygianni2023}, also exhibited relatively low temperature and high released energy. \citet{Makrygianni2023} implies that the bumps during the decay might be the signature of BFF. {\cblue{Interestingly,}} Flare II and Flare III in AT 2021aeuk display the {\cblue{post-peak}} bumps which strengthens the {\cblue{possibility of}} AT 2021aeuk as a BFF.
    \item \textbf{Spectral Features.} The lack of {\cblue{on-burst spectra}} of AT 2021aeuk {\cblue{make it hard to}} to classify it among BFF, SLSN-II, or changing-look phenomenon. However, the pre-burst spectra of AT 2021aeuk exhibit prominent BF features, specifically broad He {\sc ii}$\lambda$4686 and [N {\sc iii}]$\lambda$4640, which suggest a vigorous or potentially long-lasting process that enriches the {\cblue{local metallicity}}. The presence of BF lines, uncommon in typical optical AGN spectra \citep{vandenberk2001}, suggests the excitation by intense extreme-ultraviolet (EUV) photons around 304$\rm \AA$ \citep{osterbrock1974}. This raises the question of the origin of both the ionization source and the unusual metallicity abundance in these AGNs. Notably, nitrogen-loud quasars, identified in the UV band, are reported as a small fraction of typical AGNs \citep{Osmer1980,jianglh2008,batra2014}, and the metallicity abundance can be attributed to the local star formation in the central region \citep{collin1999, wangjm2011}, {\cblue{the existence of Wolf-Rayet stars \citep{Crowther2007}, TDE of a star going CNO cycle}} \citep{kochanek2016, liuxin2018, Leloudas2019, Blagorodnova2019, Neustadt2020, Malyali2021, mockler2024}, or {\cblue{enhancement/re-ignition of AGN accretion}} \citep{Trakhtenbrot2019, Tadhunter2017, Blanchard2017, Gromadzki2019, Frederick2019, Makrygianni2023, veres2024}. 
    \item \textbf{Multiple Flares.} Given the duration of the photometry data, it is challenging to determine whether the triple flares in AT 2021aeuk occur periodically.  Flare III exhibited a lower energy release and Eddington ratio compared to Flare II (see Table.\ref{tab_fit2}). PS1-10adi showed weak recurring behavior, which \citet{JiangN2019} attributed to shock interactions between outflows and the torus. While AT 2021vdq \citep{Somalwar2023}, F01004-2237 \citep{veres2024, SunLM2024} and AT2019aalc \citep{veres2024} exhibited recurring flares brighter than their initial outbursts, other transients, such as SN 2019meh \citep{SoraisamM2022} and AT 2021loi \citep{Makrygianni2023}, display {\cblue{sharp}} recurring flares during the decline phase. Interestingly, while the initial outburst of SN 2019meh is identified as an SLSN-II \citep{NichollM2019}, the subsequent flares exhibit sharp variations distinct from the post-peak bumps commonly explained by interactions with dense circumstellar material \citep{Hosseinzadeh2022, ChenZH2023}. Additionally, the corresponding double-peak bump in the $\emph{WISE}$ light curve of SN 2019meh (although not explicitly shown in the relevant papers) further raises questions about the true nature of SN 2019meh.
\end{itemize}

\subsection{Multiple Flares}
\label{sec_tripleflares}

While TDEs, SLSNe-II, or enhanced accretion process have been
considered as potential explanations for flares in AT 2021aeuk, the
physical mechanism responsible for the multiple occurrences remains
uncertain. The flares can be considered as independent events with
high occurrence rates under specific conditions. They can also be
considered as recurring flares induced by the related physical
process.

Galaxies with dense stellar environments, like Elliptical (E+A)
galaxies with nuclear overdensities or post-starburst galaxies
undergoing mergers, are favorable environments for a higher TDE rate
\citep{Arcavi2014, French2016, Stone2016}. Recent researches suggest
that a luminous infrared galaxy (LIRG, \citealt{Sanders1988,
Hopkins2006}), undergoing a late-stage merger event with intense
star formation activity, might experience TDEs at a higher rate
ranging from approximately $\rm (1.6-4.6)\times 10^{-3}\ yr^{-1}\
galaxy^{-1}$ \citep{Mattila2018, Kool2020, Reynolds2022}. The Subaru
image (see Fig.\ref{fig_subaru}) of J1612 reveals the minor tidal
signatures and \citet{Berton2018} report a somewhat extended radio
continuum. It also exhibits red mid-infrared color ($\emph{W1}$ -
$\emph{W2}$ = 1.1). \citet{Barrows2021} analyzed the
multi-wavelength spectral energy distribution (SED) of J1612 using
data from GALEX, SDSS DR14, Gaia DR2, PanSTARR2, and 2MASS. Their
analysis suggests a star formation rate (SFR) of approximately $6.6
\ M_{\odot}\ \rm yr^{-1}$ and a stellar mass of $ 5.4\times 10^{10}\
M_{\odot}$ after separating the AGN contribution using CIGALE. We
also give an SFR estimation of $7.0\ M_{\odot}\ \rm  yr^{-1}$ from
emission lines [O {\sc ii}] and [O {\sc iii}] \citep{ZhuangMY2020},
which is generally consistent with the SFR derived from SED fitting.
However, the broadening of the [O {\sc iii}] line introduces some
uncertainties into this measurement, and the SFR derived from SED
fitting provides a more reliable estimate. These SFR values suggest
that J1612 is unlikely to be classified as an LIRG. Notably, many
AGN-associated transients display spectral features consistent with
the NLSy1 galaxy, and J1612 was identified as an NLSy1 before the
bursts. \citet{Frederick2021} proposed that NLSy1 galaxies might be
particularly favorable environments for enhanced flare activity due
to the instabilities of high accretion rate BHs. They also point out
that it might be a selection effect for galaxies with low-mass BHs
or ignorance of the smooth flares in broad-line AGNs.

A partial TDE (pTDE), where a stellar core survives and
gravitationally interacts with the debris stream \citep{Payne2021,
Wevers2023, Malyali2023, Liuz2023, Somalwar2023}, can produce
multiple flares due to the ongoing interaction. This could be a
contributing factor to the multiple flares in AT 2021aeuk. The
models \citep{Guillochon2013, MacleodM2013, Mainetti2017,
Coughlin2019, Miles2020} also predict a steeper index $p\sim -9/4$,
which can be a possible explanation for index $p$ deviation of Flare
II in Sec.\ref{Sec_senarioofTDEs}, although we plotted the location
for a pTDE, AT 2020vdq \citep{Charalampopoulos2023, Somalwar2023,
Yao2023}, in Fig.\ref{fig_TDEsample}, and did not find much
resemblance between it and AT 2021aeuk. ASSASN-14ko is a periodic
pTDE candidate \citep{Payne2021} with a mean recurring period of
$\rm 114\pm0.4$ days. We did not examine the periodicity based on
three flares in AT 2021aeuk, however, the partial disruption reduces
the stellar mass and hence can alter the returning orbital period {\cblue{\citep{zhongsy2022}}}.
Further investigations with more pTDEs could provide valuable
insights into the possibility of the origin of AT 2021aeuk.

\citet{Chan2019, Chan2020, chan2021} carried out a series of
simulations toward the interaction between TDE and accretion disk,
where the debris stream punches through the disk, collides, and
generates shocks causing disk material inflow. The TDE dynamics and
disk mass distribution are altered, and the optical variation may
not be a simple sum of the intrinsic AGN variability and the
additional emission from the TDE. They predicted a power-law decay
index between -3 and -2 early and potentially steeper later. The
index $p$ of Flare II ($-2.99_{-0.14}^{+0.13}$) is generally
consistent with the prediction at the earlier time. They also
predicted that the luminosity can reach a plateau due to the
radiation trapping, where the inflow is faster than radiation
diffusion. Interestingly, Flare II in AT 2021aeuk exhibits a
plateau-like feature lasting for about 50 days in the g-band light
curve (see Panel a in Fig. \ref{fig_bestfit}). For J1612 with BH
mass $\rm \sim 10^{7.52}\ M_{\odot}$ and Eddington ratio $\rm \sim
0.88$ (if taking the H$\alpha$ case), it requires a disrupted star
with rather high mass. They also predicted a ``second impact'' after
the stream punched through the disk. The luminosity in X-ray or
$\gamma$-ray follows the mass-return rate but in optical/UV it does
not. Thus, this cannot sufficiently explain the subsequent Flare
III.

Another intriguing hypothesis suggests that the triple flares
observed in AT 2021aeuk might be a result of a stellar binary being
disrupted by an SMBH binary \citep{Wu2018}. The interaction between
the stellar binary and the SMBH binary can lead to multiple TDEs with varying time separations, influenced by the
BH mass ratio, star masses, etc. For a stellar binary with two stars
of 1 $M_{\odot}$ each and a semi-major axis less than 100 AU, the
time separation between flares in an SMBH binary system can range
from 150 days to 15 years. In contrast, the time separation in a
single SMBH system is generally much shorter. This model offers a
potential explanation for the observed separation of over 1000 days
between Flare II and Flare III in AT 2021aeuk. Furthermore, the
binary black hole scenario aligns with the merging features observed
in J1612, as captured by Subaru imaging.

\section{Summary}
\label{sec5}

We report a rare transient AT 2021aeuk in a radio-loud NLSy1 galaxy (SDSS J161259.82+421940.3),  with triple flares occurring between 2018 and 2023. We present archived photometry data from optical and infrared surveys, and then perform quantitative analysis based on the optical light curves and conduct spectral decomposition on the pre-burst spectra. The results are as follows,\\

\begin{itemize}
    \item Despite the strong degeneracy between $p$ and $t_0$, the power-law decay index of Flare II ($p=-2.99_{-0.14}^{+0.13}$) is rather smooth compared to the current TDE sample, while Flare III's ($-1.61_{-0.65}^{+0.34}$) is generally consistent with the theoretical $-5/3$.\\
    \item The blackbody temperature shows minor evolution of $\sim 10^{3.8}$ K, much lower than the TDE sample. The $t_{1/2}$, blackbody radius {\cblue{($\gtrsim 10^{16}\ cm$)}} and released energy $\sim 10^{52}\rm\ erg$ are generally larger.\\
    \item The time lag ($3.4_{-0.9}^{+1.0}$ days) between $g$- and $r$-bands is significantly larger ($\sim 12.1$ times) than that from standard accretion disk prediction, and the peak variation of $g-$ and $r-$ band displays a different pattern. This indicates that the flares cannot solely be attributed to the typical AGN accretion process.\\
    \item {\cblue{The time lag ($103.6_{-38.7}^{+71.7}$) between $g$- and $W1$-bands is significantly smaller than the value from $R-L$ prediction ($\sim 325$ days). However, the sampling deficiency of the mid-infrared light curve introduces some uncertainties.}}\\
    \item The two pre-burst spectra show BF lines of [N {\sc iii}]$\lambda$4640 indicating a vigorous or potentially long-lasting process that enriches the local metallicity. We derived a BH mass  $\rm logM_{\bullet}=7.09^{+0.18}_{-0.31}\ M_{\odot}$ and $\rm logM_{\bullet}=7.52^{+0.08}_{-0.10}\ M_{\odot}$ using the FWHM of $\rm H\beta$ and $\rm H\alpha$ from spectral fitting, respectively, which are generally consistent with previous measurements. {\cblue{The corresponding AGN accretion rate is $\sim1.74$ and $\sim0.88$.}}\\
    \item {\cblue{The Eddington ratio at peak luminosity of Flare II and III are $\sim 0.58$ and $\sim 0.38$.}} \\
\end{itemize}

We compare AT 2021aeuk with blazars, TDEs, typical AGN intrinsic
variation, SLSNe-II, and ambiguous flares. The possibility of
blazars, typical AGN intrinsic variation {\cblue{, and SLSNe-II can
be excluded based on the variation patterns, time delay, and recurring flares.}} AT 2021aeuk exhibited similar evolving patterns
to TDEs. {\cblue{However,}} the long-term decay and the deviating blackbody parameters raise questions about TDE origin. {\cblue{The unconventional behaviors may induced by the coupling process between transients and accretion disk.}} The multiple
flares can be explained as independent events {\cblue{with high occurrence}} in certain
environments. It can also be explained as recurring flares
{\cblue{induced by}} the related physical processes.
Multi-wavelength follow-up observations will be beneficial in
determining the underlying physical mechanism responsible for AT
2021aeuk.

\section*{acknowledgements}
We really appreciate the referee for
many precious comments that significantly improved the manuscript.
Dong-Wei Bao and Wei-Jian Guo contributed equally to this work. We
thank  Hong Wu for using the Xinglong 2.16 m telescope to take a
tentative spectrum of J1612 and Yu-Yang Songsheng for valuable
discussions. {\cblue{We thank James Wicker for the invaluable help with polishing the language and correcting the language issues.}} We thank for Jun-Rong Liu's $\it Fermi$-LAT data
reduction and useful talk. We thank for Heng-Xiao Guo's precious
communication. This work was supported by National Science
Foundation of China No. 11988101, 11973051. We acknowledge the
support from the National Key R\&D Program of China (2020YFC2201400,
2021YFA1600404), National Natural Science Foundation of China
(NSFC-11991050, -11991054, -11833008, -12103041, -12263003,
-12333003), the International Partnership Program of the Chinese
Academy of Sciences (113111KYSB20200014) and the science research
grants from the China Manned Space Project with NO.
CMS-CSST-2021-A05. W. Guo and H. Zou acknowledge the supports from
National Key R\&D Program of China (grant Nos. 2023YFA1607800,
2022YFA1602902, 2023YFA1608100, 2023YFF0714800, and 2023YFA1608303),
the National Natural Science Foundation of China (NSFC; grant Nos.
12120101003, 12373010, 12233008, and 12173051), Beijing Municipal
Natural Science Foundation (grant No. 1222028), and the science
research grants from the China Manned Space Project with Nos.
CMS-CSST-2021-A02 and CMS-CSST-2021-A04. YFY is supported by the
Strategic Priority Research Program of the Chinese Academy of
Sciences, Grant No. XDB0550200 and National SKA Program of China No.
2020SKA0120300. YRL acknowledges financial support from the NSFC
through grant No 12273041 and from the Youth Innovation Promotion
Associatin CAS.  C.C. is supported by the National Natural Science
Foundation of China, No. 11803044, 11933003, 12173045. This work is
sponsored (in part) by the Chinese Academy of Sciences (CAS),
through a grant to the CAS South America Center for Astronomy
(CASSACA).

We also acknowledge SDSS for providing extensive spectral database
support. SDSS is managed by the Astrophysical Research Consortium
for the Participating Institutions of the SDSS Collaboration
including the Brazilian Participation Group, the Carnegie
Institution for Science, Carnegie Mellon University, Center for
Astrophysics | Harvard \& Smithsonian, the Chilean Participation
Group, the French Participation Group, Instituto de Astrofísica de
Canarias, The Johns Hopkins University, Kavli Institute for the
Physics and Mathematics of the Universe(IPMU)/University of Tokyo,
the Korean Participation Group, Lawrence Berkeley National
Laboratory, Leibniz Institut fürAstrophysik Potsdam (AIP),
Max-Planck-Institut für Astronomie (MPIA Heidelberg),
Max-Planck-Institut für Astrophysik (MPA Garching),
Max-Planck-Institut für ExtraterrestrischePhysik (MPE), National
Astronomical Observatories of China, New Mexico State University,
New York University, University of Notre Dame, Observatário
Nacional/MCTI, The Ohio State University, Pennsylvania State
University, Shanghai Astronomical Observatory, United Kingdom
Participation Group, Universidad Nacional Autónoma de México,
University of Arizona, University of Colorado Boulder, University of
Oxford, University of Portsmouth, University of Utah, University of
Virginia, University of Washington, University of Wisconsin,
Vanderbilt University, and Yale University.

We acknowledge the efforts for public data from CTRS, PS1, ASAS-SN
and ZTF. The Catalina Sky Survey is funded by the National
Aeronautics and Space Administration under Grant No. NNG05GF22G
issued through the Science Mission Directorate Near-Earth Objects
Observations Program. The CRTS survey is supported by the US
National Science Foundation under grants AST-0909182 and
AST-1313422. The CRTS survey is supported by the US National Science
Foundation under grants AST-0909182 and AST-1313422.

The PS1 has been made possible through contributions by the
Institute for Astronomy, the University of Hawaii, the Pan-STARRS
Project Office, the Max-Planck Society and its participating in-
stitutes, the Max Planck Institute for Astronomy, Heidelberg and the
Max Planck Institute for Extraterrestrial Physics, Garching, The
Johns Hopkins University, Durham University, the University of
Edinburgh, Queen’s University Belfast, the Harvard-Smithsonian
Center for Astrophysics, the Las Cumbres Observatory Global
Telescope Network Incorporated, the National Central University of
Taiwan, the Space Telescope Science Institute, the National
Aeronautics and Space Administration under Grant No. NNX08AR22G
issued through the Planetary Science Division of the NASA Science
Mission Directorate, the National Science Foundation under Grant No.
AST-1238877, the University of Maryland, and Eotvos Lorand
University (ELTE).

ASAS-SN is supported by the Gordon and Betty Moore Foundation
through grant GBMF5490 to the Ohio State University and NSF grant
AST-1515927. Development of ASAS-SN has been supported by NSF grant
AST-0908816, the Mt. Cuba Astronomical Foundation, the Center for
Cosmology and Astro- Particle Physics at the Ohio State University,
the Chinese Academy of Sciences South America Center for Astronomy
(CASSACA), the Villum Foundation, and George Skestos.

ZTF is supported by the National Science Foundation under Grant No.
AST-2034437 and a collaboration including Caltech, IPAC, the
Weizmann Institute for Science, the Oskar Klein Center at Stockholm
University, the University of Maryland, Deutsches
Elektronen-Synchrotron and Humboldt University, the TANGO Consortium
of Taiwan, the University of Wis- consin at Milwaukee, Trinity
College Dublin, Lawrence Livermore National Laboratories, and IN2P3,
France. Operations are conducted by COO, IPAC, and UW.

This publication makes use of data products from $\emph{WISE}$ ,
which is a joint project of the University of California, Los
Angeles, and the Jet Propulsion Laboratory/California Institute of
Technology, funded by the National Aeronautics and Space
Administration. This publication also makes use of data products
from $NEOWISE$, which is a project of the Jet Propulsion
Laboratory/California Institute of Technology, funded by the
Planetary Science Division of the National Aeronautics and Space
Administration.

\clearpage
\appendix

\setcounter{figure}{0}
\section{Appendix Figure}

\renewcommand{\thefigure}{A\arabic{figure}}
\begin{figure}[H]
    \centering
    \includegraphics[width=0.85\textwidth]{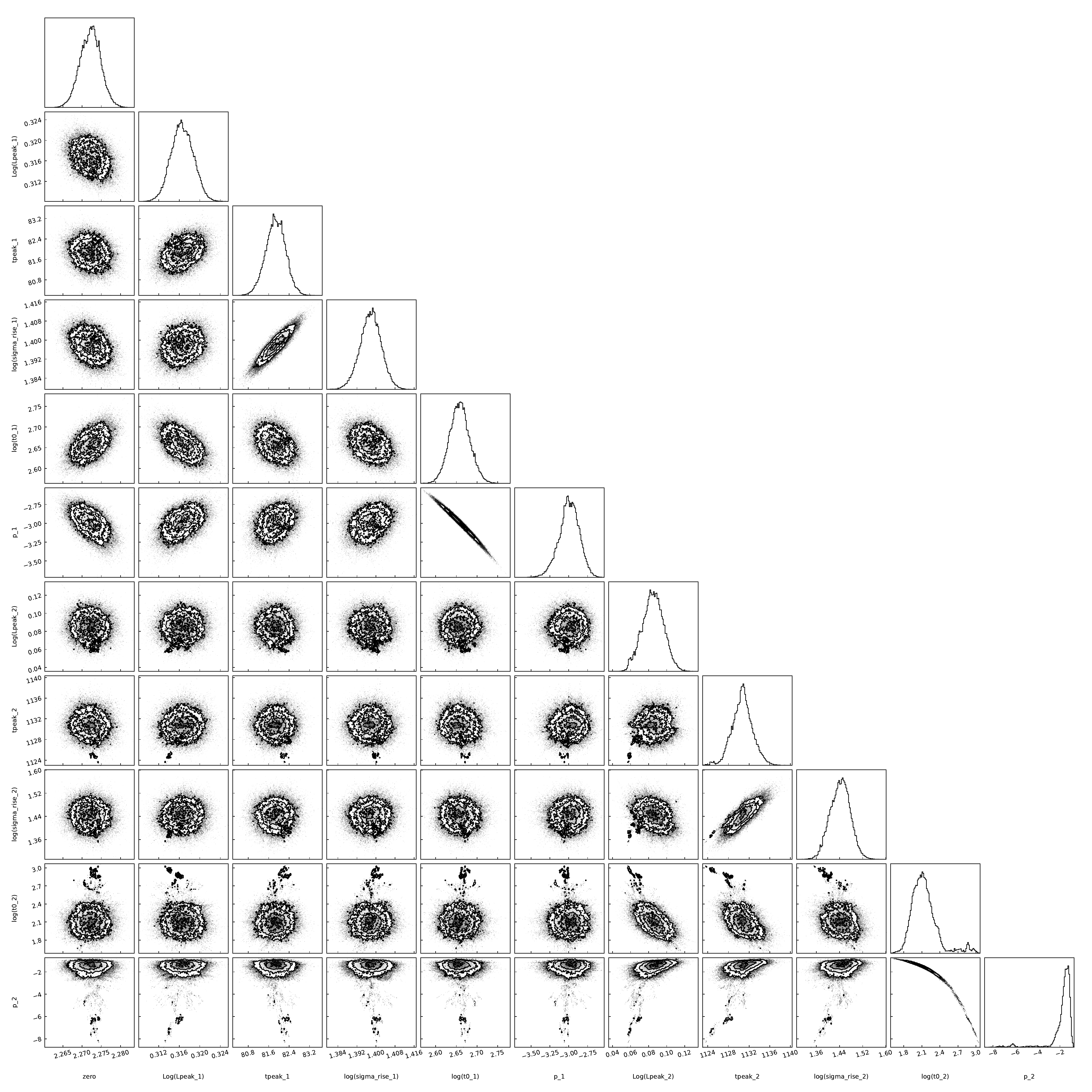}
    \caption{
    The MCMC posterior distributions of $L_0$, $L_{\nu}^{\rm peak}$, $t_{\rm peak}$, $\sigma$, $t_{0}$ and $p$ for Flare II and Flare III in Sec.\ref{sec_lcfit}. The contours are at 1$\sigma$, 1.5$\sigma$, and 2$\sigma$ levels. Note that $t_{\rm peak}$, $\sigma$ and $t_{0}$ are in the rest frame.}
    \label{fig_lccornerplot}
\end{figure}

\bibliographystyle{aasjournal}
\bibliography{reference}

\begin{thebibliography}{}
\expandafter\ifx\csname natexlab\endcsname\relax\def\natexlab#1{#1}\fi
\providecommand{\url}[1]{\href{#1}{#1}}
\providecommand{\dodoi}[1]{doi:~\href{http://doi.org/#1}{\nolinkurl{#1}}}
\providecommand{\doeprint}[1]{\href{http://ascl.net/#1}{\nolinkurl{http://ascl.net/#1}}}
\providecommand{\doarXiv}[1]{\href{https://arxiv.org/abs/#1}{\nolinkurl{https://arxiv.org/abs/#1}}}

\bibitem[{{Ai} {et~al.}(2010){Ai}, {Yuan}, {Zhou}, {Wang}, {Dong}, {Wang}, \& {Lu}}]{AiYL2010}
{Ai}, Y.~L., {Yuan}, W., {Zhou}, H.~Y., {et~al.} 2010, \apjl, 716, L31, \dodoi{10.1088/2041-8205/716/1/L31}

\bibitem[{{Aihara} {et~al.}(2022){Aihara}, {AlSayyad}, {Ando}, {Armstrong}, {Bosch}, {Egami}, {Furusawa}, {Furusawa}, {Harasawa}, {Harikane}, {Hsieh}, {Ikeda}, {Ito}, {Iwata}, {Kodama}, {Koike}, {Kokubo}, {Komiyama}, {Li}, {Liang}, {Lin}, {Lupton}, {Lust}, {MacArthur}, {Mawatari}, {Mineo}, {Miyatake}, {Miyazaki}, {More}, {Morishima}, {Murayama}, {Nakajima}, {Nakata}, {Nishizawa}, {Oguri}, {Okabe}, {Okura}, {Ono}, {Osato}, {Ouchi}, {Pan}, {Plazas Malag{\'o}n}, {Price}, {Reed}, {Rykoff}, {Shibuya}, {Simunovic}, {Strauss}, {Sugimori}, {Suto}, {Suzuki}, {Takada}, {Takagi}, {Takata}, {Takita}, {Tanaka}, {Tang}, {Taranu}, {Terai}, {Toba}, {Turner}, {Uchiyama}, {Vijarnwannaluk}, {Waters}, {Yamada}, {Yamamoto}, \& {Yamashita}}]{2022PASJ...74..247A}
{Aihara}, H., {AlSayyad}, Y., {Ando}, M., {et~al.} 2022, \pasj, 74, 247, \dodoi{10.1093/pasj/psab122}

\bibitem[{{Araudo} {et~al.}(2010){Araudo}, {Bosch-Ramon}, \& {Romero}}]{Araudo2010}
{Araudo}, A.~T., {Bosch-Ramon}, V., \& {Romero}, G.~E. 2010, \aap, 522, A97, \dodoi{10.1051/0004-6361/201014660}

\bibitem[{{Arcavi} {et~al.}(2014){Arcavi}, {Gal-Yam}, {Sullivan}, {Pan}, {Cenko}, {Horesh}, {Ofek}, {De Cia}, {Yan}, {Yang}, {Howell}, {Tal}, {Kulkarni}, {Tendulkar}, {Tang}, {Xu}, {Sternberg}, {Cohen}, {Bloom}, {Nugent}, {Kasliwal}, {Perley}, {Quimby}, {Miller}, {Theissen}, \& {Laher}}]{Arcavi2014}
{Arcavi}, I., {Gal-Yam}, A., {Sullivan}, M., {et~al.} 2014, \apj, 793, 38, \dodoi{10.1088/0004-637X/793/1/38}

\bibitem[{{Balbus} \& {Hawley}(1991)}]{Balbus1991}
{Balbus}, S.~A., \& {Hawley}, J.~F. 1991, \apj, 376, 214, \dodoi{10.1086/170270}

\bibitem[{{Barrows} {et~al.}(2021){Barrows}, {Comerford}, {Stern}, \& {Assef}}]{Barrows2021}
{Barrows}, R.~S., {Comerford}, J.~M., {Stern}, D., \& {Assef}, R.~J. 2021, \apj, 922, 179, \dodoi{10.3847/1538-4357/ac1352}

\bibitem[{{Barvainis}(1987)}]{Barvainis1987}
{Barvainis}, R. 1987, \apj, 320, 537, \dodoi{10.1086/165571}

\bibitem[{{Batra} \& {Baldwin}(2014)}]{batra2014}
{Batra}, N.~D., \& {Baldwin}, J.~A. 2014, \mnras, 439, 771, \dodoi{10.1093/mnras/stu007}

\bibitem[{{Bentz} {et~al.}(2013){Bentz}, {Denney}, {Grier}, {Barth}, {Peterson}, {Vestergaard}, {Bennert}, {Canalizo}, {De Rosa}, {Filippenko}, {Gates}, {Greene}, {Li}, {Malkan}, {Pogge}, {Stern}, {Treu}, \& {Woo}}]{Bentz2013}
{Bentz}, M.~C., {Denney}, K.~D., {Grier}, C.~J., {et~al.} 2013, \apj, 767, 149, \dodoi{10.1088/0004-637X/767/2/149}

\bibitem[{{Berton} {et~al.}(2018){Berton}, {Congiu}, {J{\"a}rvel{\"a}}, {Antonucci}, {Kharb}, {Lister}, {Tarchi}, {Caccianiga}, {Chen}, {Foschini}, {L{\"a}hteenm{\"a}ki}, {Richards}, {Ciroi}, {Cracco}, {Frezzato}, {La Mura}, \& {Rafanelli}}]{Berton2018}
{Berton}, M., {Congiu}, E., {J{\"a}rvel{\"a}}, E., {et~al.} 2018, \aap, 614, A87, \dodoi{10.1051/0004-6361/201832612}

\bibitem[{{Blagorodnova} {et~al.}(2019){Blagorodnova}, {Cenko}, {Kulkarni}, {Arcavi}, {Bloom}, {Duggan}, {Filippenko}, {Fremling}, {Horesh}, {Hosseinzadeh}, {Karamehmetoglu}, {Levan}, {Masci}, {Nugent}, {Pasham}, {Veilleux}, {Walters}, {Yan}, \& {Zheng}}]{Blagorodnova2019}
{Blagorodnova}, N., {Cenko}, S.~B., {Kulkarni}, S.~R., {et~al.} 2019, \apj, 873, 92, \dodoi{10.3847/1538-4357/ab04b0}

\bibitem[{{Blanchard} {et~al.}(2017){Blanchard}, {Nicholl}, {Berger}, {Guillochon}, {Margutti}, {Chornock}, {Alexander}, {Leja}, \& {Drout}}]{Blanchard2017}
{Blanchard}, P.~K., {Nicholl}, M., {Berger}, E., {et~al.} 2017, \apj, 843, 106, \dodoi{10.3847/1538-4357/aa77f7}

\bibitem[{{Blandford} \& {McKee}(1982)}]{Blandford1982}
{Blandford}, R.~D., \& {McKee}, C.~F. 1982, \apj, 255, 419, \dodoi{10.1086/159843}

\bibitem[{{Bonning} {et~al.}(2012){Bonning}, {Urry}, {Bailyn}, {Buxton}, {Chatterjee}, {Coppi}, {Fossati}, {Isler}, \& {Maraschi}}]{Bonning2012}
{Bonning}, E., {Urry}, C.~M., {Bailyn}, C., {et~al.} 2012, \apj, 756, 13, \dodoi{10.1088/0004-637X/756/1/13}

\bibitem[{{B{\"o}ttcher}(2019)}]{Bottcher2019}
{B{\"o}ttcher}, M. 2019, Galaxies, 7, 20, \dodoi{10.3390/galaxies7010020}

\bibitem[{{Bowen}(1928)}]{Bowen1928}
{Bowen}, I.~S. 1928, \apj, 67, 1, \dodoi{10.1086/143091}

\bibitem[{{Cackett} {et~al.}(2021){Cackett}, {Bentz}, \& {Kara}}]{Cackett2021}
{Cackett}, E.~M., {Bentz}, M.~C., \& {Kara}, E. 2021, iScience, 24, 102557, \dodoi{10.1016/j.isci.2021.102557}

\bibitem[{{Cackett} {et~al.}(2007){Cackett}, {Horne}, \& {Winkler}}]{Cackett2007}
{Cackett}, E.~M., {Horne}, K., \& {Winkler}, H. 2007, \mnras, 380, 669, \dodoi{10.1111/j.1365-2966.2007.12098.x}

\bibitem[{{Cannizzaro} {et~al.}(2022){Cannizzaro}, {Levan}, {van Velzen}, \& {Brown}}]{Cannizzaro2022}
{Cannizzaro}, G., {Levan}, A.~J., {van Velzen}, S., \& {Brown}, G. 2022, \mnras, 516, 529, \dodoi{10.1093/mnras/stac2014}

\bibitem[{{Chambers} {et~al.}(2016){Chambers}, {Magnier}, {Metcalfe}, {Flewelling}, {Huber}, {Waters}, {Denneau}, {Draper}, {Farrow}, {Finkbeiner}, {Holmberg}, {Koppenhoefer}, {Price}, {Rest}, {Saglia}, {Schlafly}, {Smartt}, {Sweeney}, {Wainscoat}, {Burgett}, {Chastel}, {Grav}, {Heasley}, {Hodapp}, {Jedicke}, {Kaiser}, {Kudritzki}, {Luppino}, {Lupton}, {Monet}, {Morgan}, {Onaka}, {Shiao}, {Stubbs}, {Tonry}, {White}, {Ba{\~n}ados}, {Bell}, {Bender}, {Bernard}, {Boegner}, {Boffi}, {Botticella}, {Calamida}, {Casertano}, {Chen}, {Chen}, {Cole}, {Deacon}, {Frenk}, {Fitzsimmons}, {Gezari}, {Gibbs}, {Goessl}, {Goggia}, {Gourgue}, {Goldman}, {Grant}, {Grebel}, {Hambly}, {Hasinger}, {Heavens}, {Heckman}, {Henderson}, {Henning}, {Holman}, {Hopp}, {Ip}, {Isani}, {Jackson}, {Keyes}, {Koekemoer}, {Kotak}, {Le}, {Liska}, {Long}, {Lucey}, {Liu}, {Martin}, {Masci}, {McLean}, {Mindel}, {Misra}, {Morganson}, {Murphy}, {Obaika}, {Narayan}, {Nieto-Santisteban}, {Norberg}, {Peacock}, {Pier}, {Postman}, {Primak}, {Rae}, {Rai},
  {Riess}, {Riffeser}, {Rix}, {R{\"o}ser}, {Russel}, {Rutz}, {Schilbach}, {Schultz}, {Scolnic}, {Strolger}, {Szalay}, {Seitz}, {Small}, {Smith}, {Soderblom}, {Taylor}, {Thomson}, {Taylor}, {Thakar}, {Thiel}, {Thilker}, {Unger}, {Urata}, {Valenti}, {Wagner}, {Walder}, {Walter}, {Watters}, {Werner}, {Wood-Vasey}, \& {Wyse}}]{Chambers2016}
{Chambers}, K.~C., {Magnier}, E.~A., {Metcalfe}, N., {et~al.} 2016, arXiv e-prints, arXiv:1612.05560, \dodoi{10.48550/arXiv.1612.05560}

\bibitem[{{Chan} {et~al.}(2020){Chan}, {Piran}, \& {Krolik}}]{Chan2020}
{Chan}, C.-H., {Piran}, T., \& {Krolik}, J.~H. 2020, \apj, 903, 17, \dodoi{10.3847/1538-4357/abb776}

\bibitem[{{Chan} {et~al.}(2021){Chan}, {Piran}, \& {Krolik}}]{chan2021}
---. 2021, \apj, 914, 107, \dodoi{10.3847/1538-4357/abf0a7}

\bibitem[{{Chan} {et~al.}(2019){Chan}, {Piran}, {Krolik}, \& {Saban}}]{Chan2019}
{Chan}, C.-H., {Piran}, T., {Krolik}, J.~H., \& {Saban}, D. 2019, \apj, 881, 113, \dodoi{10.3847/1538-4357/ab2b40}

\bibitem[{{Chand} \& {Gopal-Krishna}(2022)}]{Chand2022}
{Chand}, K., \& {Gopal-Krishna}. 2022, \mnras, 516, L18, \dodoi{10.1093/mnrasl/slac066}

\bibitem[{{Charalampopoulos} {et~al.}(2023){Charalampopoulos}, {Leloudas}, {Pursiainen}, \& {Kotak}}]{Charalampopoulos2023}
{Charalampopoulos}, P., {Leloudas}, G., {Pursiainen}, M., \& {Kotak}, R. 2023, Transient Name Server AstroNote, 115, 1

\bibitem[{{Chen} {et~al.}(2023{\natexlab{a}}){Chen}, {Liu}, {Zhai}, {Yao}, {Li}, {Du}, {Hu}, {Guo}, {Xiao}, {Songsheng}, \& {Wang}}]{chen2023}
{Chen}, Y.-J., {Liu}, J.-R., {Zhai}, S., {et~al.} 2023{\natexlab{a}}, \mnras, 522, 3439, \dodoi{10.1093/mnras/stad1136}

\bibitem[{{Chen} {et~al.}(2023{\natexlab{b}}){Chen}, {Yan}, {Kangas}, {Lunnan}, {Sollerman}, {Schulze}, {Perley}, {Chen}, {Taggart}, {Hinds}, {Gal-Yam}, {Wang}, {De}, {Bellm}, {Bloom}, {Dekany}, {Graham}, {Kasliwal}, {Kulkarni}, {Laher}, {Neill}, \& {Rusholme}}]{ChenZH2023}
{Chen}, Z.~H., {Yan}, L., {Kangas}, T., {et~al.} 2023{\natexlab{b}}, \apj, 943, 42, \dodoi{10.3847/1538-4357/aca162}

\bibitem[{{Collin} \& {Zahn}(1999)}]{collin1999}
{Collin}, S., \& {Zahn}, J.-P. 1999, \aap, 344, 433

\bibitem[{{Coughlin} \& {Nixon}(2019)}]{Coughlin2019}
{Coughlin}, E.~R., \& {Nixon}, C.~J. 2019, \apjl, 883, L17, \dodoi{10.3847/2041-8213/ab412d}

\bibitem[{{Cracco} {et~al.}(2016){Cracco}, {Ciroi}, {Berton}, {Di Mille}, {Foschini}, {La Mura}, \& {Rafanelli}}]{CraccoV2016}
{Cracco}, V., {Ciroi}, S., {Berton}, M., {et~al.} 2016, \mnras, 462, 1256, \dodoi{10.1093/mnras/stw1689}

\bibitem[{{Crowther}(2007)}]{Crowther2007}
{Crowther}, P.~A. 2007, \araa, 45, 177, \dodoi{10.1146/annurev.astro.45.051806.110615}

\bibitem[{{Dgany} {et~al.}(2023){Dgany}, {Arcavi}, {Makrygianni}, {Pellegrino}, \& {Howell}}]{Dgany2023}
{Dgany}, Y., {Arcavi}, I., {Makrygianni}, L., {Pellegrino}, C., \& {Howell}, D.~A. 2023, \apj, 957, 57, \dodoi{10.3847/1538-4357/ace971}

\bibitem[{{Dong} {et~al.}(2016){Dong}, {Chen}, {Bose}, {Stanek}, {Kochanek}, {Holoien}, {Shappee}, {Prieto}, {Brown}, \& {Milne}}]{DongS2016}
{Dong}, S., {Chen}, P., {Bose}, S., {et~al.} 2016, The Astronomer's Telegram, 9843, 1

\bibitem[{{Dou} {et~al.}(2017){Dou}, {Wang}, {Yan}, {Jiang}, {Yang}, {Cutri}, {Mainzer}, \& {Peng}}]{DouLM2017}
{Dou}, L., {Wang}, T., {Yan}, L., {et~al.} 2017, \apjl, 841, L8, \dodoi{10.3847/2041-8213/aa7130}

\bibitem[{{Drake} {et~al.}(2009){Drake}, {Djorgovski}, {Mahabal}, {Beshore}, {Larson}, {Graham}, {Williams}, {Christensen}, {Catelan}, {Boattini}, {Gibbs}, {Hill}, \& {Kowalski}}]{Drake2009}
{Drake}, A.~J., {Djorgovski}, S.~G., {Mahabal}, A., {et~al.} 2009, \apj, 696, 870, \dodoi{10.1088/0004-637X/696/1/870}

\bibitem[{{Drake} {et~al.}(2011){Drake}, {Djorgovski}, {Mahabal}, {Anderson}, {Roy}, {Mohan}, {Ravindranath}, {Frail}, {Gezari}, {Neill}, {Ho}, {Prieto}, {Thompson}, {Thorstensen}, {Wagner}, {Kowalski}, {Chiang}, {Grove}, {Schinzel}, {Wood}, {Carrasco}, {Recillas}, {Kewley}, {Archana}, {Basu}, {Wadadekar}, {Kumar}, {Myers}, {Phinney}, {Williams}, {Graham}, {Catelan}, {Beshore}, {Larson}, \& {Christensen}}]{DrakeAJ2011}
---. 2011, \apj, 735, 106, \dodoi{10.1088/0004-637X/735/2/106}

\bibitem[{{Du} {et~al.}(2016){Du}, {Lu}, {Zhang}, {Huang}, {Wang}, {Hu}, {Qiu}, {Li}, {Fan}, {Fang}, {Bai}, {Bian}, {Yuan}, {Ho}, {Wang}, \& {SEAMBH Collaboration}}]{DuPu2016}
{Du}, P., {Lu}, K.-X., {Zhang}, Z.-X., {et~al.} 2016, \apj, 825, 126, \dodoi{10.3847/0004-637X/825/2/126}

\bibitem[{{Evans} \& {Kochanek}(1989)}]{Evans1989}
{Evans}, C.~R., \& {Kochanek}, C.~S. 1989, \apjl, 346, L13, \dodoi{10.1086/185567}

\bibitem[{{Fausnaugh} {et~al.}(2018){Fausnaugh}, {Starkey}, {Horne}, {Kochanek}, {Peterson}, {Bentz}, {Denney}, {Grier}, {Grupe}, {Pogge}, {De Rosa}, {Adams}, {Barth}, {Beatty}, {Bhattacharjee}, {Borman}, {Boroson}, {Bottorff}, {Brown}, {Brown}, {Brotherton}, {Coker}, {Crawford}, {Croxall}, {Eftekharzadeh}, {Eracleous}, {Joner}, {Henderson}, {Holoien}, {Hutchison}, {Kaspi}, {Kim}, {King}, {Li}, {Lochhaas}, {Ma}, {MacInnis}, {Manne-Nicholas}, {Mason}, {Montuori}, {Mosquera}, {Mudd}, {Musso}, {Nazarov}, {Nguyen}, {Okhmat}, {Onken}, {Ou-Yang}, {Pancoast}, {Pei}, {Penny}, {Poleski}, {Rafter}, {Romero-Colmenero}, {Runnoe}, {Sand}, {Schimoia}, {Sergeev}, {Shappee}, {Simonian}, {Somers}, {Spencer}, {Stevens}, {Tayar}, {Treu}, {Valenti}, {Van Saders}, {Villanueva}, {Villforth}, {Weiss}, {Winkler}, \& {Zhu}}]{Fausnaugh2018}
{Fausnaugh}, M.~M., {Starkey}, D.~A., {Horne}, K., {et~al.} 2018, \apj, 854, 107, \dodoi{10.3847/1538-4357/aaaa2b}

\bibitem[{{Fitzpatrick}(1999)}]{Fitzpatrick1999PASP}
{Fitzpatrick}, E.~L. 1999, PASP, 111, 63, \dodoi{10.1086/316293}

\bibitem[{{Foreman-Mackey} {et~al.}(2013){Foreman-Mackey}, {Hogg}, {Lang}, \& {Goodman}}]{Foreman-Mackey2013}
{Foreman-Mackey}, D., {Hogg}, D.~W., {Lang}, D., \& {Goodman}, J. 2013, \pasp, 125, 306, \dodoi{10.1086/670067}

\bibitem[{{Frank} {et~al.}(2002){Frank}, {King}, \& {Raine}}]{FrankJ2002}
{Frank}, J., {King}, A., \& {Raine}, D.~J. 2002, {Accretion Power in Astrophysics: Third Edition}

\bibitem[{{Frederick} {et~al.}(2019){Frederick}, {Gezari}, {Graham}, {Cenko}, {van Velzen}, {Stern}, {Blagorodnova}, {Kulkarni}, {Yan}, {De}, {Fremling}, {Hung}, {Kara}, {Shupe}, {Ward}, {Bellm}, {Dekany}, {Duev}, {Feindt}, {Giomi}, {Kupfer}, {Laher}, {Masci}, {Miller}, {Neill}, {Ngeow}, {Patterson}, {Porter}, {Rusholme}, {Sollerman}, \& {Walters}}]{Frederick2019}
{Frederick}, S., {Gezari}, S., {Graham}, M.~J., {et~al.} 2019, \apj, 883, 31, \dodoi{10.3847/1538-4357/ab3a38}

\bibitem[{{Frederick} {et~al.}(2021){Frederick}, {Gezari}, {Graham}, {Sollerman}, {van Velzen}, {Perley}, {Stern}, {Ward}, {Hammerstein}, {Hung}, {Yan}, {Andreoni}, {Bellm}, {Duev}, {Kowalski}, {Mahabal}, {Masci}, {Medford}, {Rusholme}, {Smith}, \& {Walters}}]{Frederick2021}
---. 2021, \apj, 920, 56, \dodoi{10.3847/1538-4357/ac110f}

\bibitem[{{French} {et~al.}(2016){French}, {Arcavi}, \& {Zabludoff}}]{French2016}
{French}, K.~D., {Arcavi}, I., \& {Zabludoff}, A. 2016, \apjl, 818, L21, \dodoi{10.3847/2041-8205/818/1/L21}

\bibitem[{{Gaskell} {et~al.}(1986){Gaskell}, {Cappellaro}, {Dinerstein}, {Garnett}, {Harkness}, \& {Wheeler}}]{Gaskell1986}
{Gaskell}, C.~M., {Cappellaro}, E., {Dinerstein}, H.~L., {et~al.} 1986, \apjl, 306, L77, \dodoi{10.1086/184709}

\bibitem[{{Gezari}(2012)}]{Gezari2012}
{Gezari}, S. 2012, in European Physical Journal Web of Conferences, Vol.~39, European Physical Journal Web of Conferences, 03001, \dodoi{10.1051/epjconf/20123903001}

\bibitem[{{Gezari}(2021)}]{Gezari2021}
{Gezari}, S. 2021, \araa, 59, 21, \dodoi{10.1146/annurev-astro-111720-030029}

\bibitem[{{Gezari} {et~al.}(2009){Gezari}, {Halpern}, {Grupe}, {Yuan}, {Quimby}, {McKay}, {Chamarro}, {Sisson}, {Akerlof}, {Wheeler}, {Brown}, {Cenko}, {Rau}, {Djordjevic}, \& {Terndrup}}]{GezariS2009}
{Gezari}, S., {Halpern}, J.~P., {Grupe}, D., {et~al.} 2009, \apj, 690, 1313, \dodoi{10.1088/0004-637X/690/2/1313}

\bibitem[{{Gopal-Krishna} {et~al.}(2011){Gopal-Krishna}, {Goyal}, {Joshi}, {Karthick}, {Sagar}, {Wiita}, {Anupama}, \& {Sahu}}]{Gopal2011}
{Gopal-Krishna}, {Goyal}, A., {Joshi}, S., {et~al.} 2011, \mnras, 416, 101, \dodoi{10.1111/j.1365-2966.2011.19014.x}

\bibitem[{{Goyal} {et~al.}(2018){Goyal}, {Stawarz}, {Zola}, {Marchenko}, {Soida}, {Nilsson}, {Ciprini}, {Baran}, {Ostrowski}, {Wiita}, {Gopal-Krishna}, {Siemiginowska}, {Sobolewska}, {Jorstad}, {Marscher}, {Aller}, {Aller}, {Hovatta}, {Caton}, {Reichart}, {Matsumoto}, {Sadakane}, {Gazeas}, {Kidger}, {Piirola}, {Jermak}, {Alicavus}, {Baliyan}, {Baransky}, {Berdyugin}, {Blay}, {Boumis}, {Boyd}, {Bufan}, {Campas Torrent}, {Campos}, {Carrillo G{\'o}mez}, {Dalessio}, {Debski}, {Dimitrov}, {Drozdz}, {Er}, {Erdem}, {Escartin P{\'e}rez}, {Fallah Ramazani}, {Filippenko}, {Gafton}, {Garcia}, {Godunova}, {G{\'o}mez Pinilla}, {Gopinathan}, {Haislip}, {Haque}, {Harmanen}, {Hudec}, {Hurst}, {Ivarsen}, {Joshi}, {Kagitani}, {Karaman}, {Karjalainen}, {Kaur}, {Kozie{\l}-Wierzbowska}, {Kuligowska}, {Kundera}, {Kurowski}, {Kvammen}, {LaCluyze}, {Lee}, {Liakos}, {Lozano de Haro}, {Moore}, {Mugrauer}, {Naves Nogues}, {Neely}, {Ogloza}, {Okano}, {Pajdosz}, {Pandey}, {Perri}, {Poyner}, {Provencal}, {Pursimo}, {Raj}, {Rajkumar},
  {Reinthal}, {Reynolds}, {Saario}, {Sadegi}, {Sakanoi}, {Salto Gonz{\'a}lez}, {Sameer}, {Simon}, {Siwak}, {Schweyer}, {Sold{\'a}n Alfaro}, {Sonbas}, {Strobl}, {Takalo}, {Tremosa Espasa}, {Valdes}, {Vasylenko}, {Verrecchia}, {Webb}, {Yoneda}, {Zejmo}, {Zheng}, {Zielinski}, {Janik}, {Chavushyan}, {Mohammed}, {Cheung}, \& {Giroletti}}]{Goyal2018}
{Goyal}, A., {Stawarz}, {\L}., {Zola}, S., {et~al.} 2018, \apj, 863, 175, \dodoi{10.3847/1538-4357/aad2de}

\bibitem[{{Green} {et~al.}(2022){Green}, {Pulgarin-Duque}, {Anderson}, {MacLeod}, {Eracleous}, {Ruan}, {Runnoe}, {Graham}, {Roulston}, {Schneider}, {Ahlf}, {Bizyaev}, {Brownstein}, {del Casal}, {Dodd}, {Hoover}, {Matt}, {Merloni}, {Pan}, {Ramirez}, {Ridder}, \& {Moseley}}]{GreenPJ2022}
{Green}, P.~J., {Pulgarin-Duque}, L., {Anderson}, S.~F., {et~al.} 2022, \apj, 933, 180, \dodoi{10.3847/1538-4357/ac743f}

\bibitem[{{Greene} \& {Ho}(2007)}]{GreeneJ2007}
{Greene}, J.~E., \& {Ho}, L.~C. 2007, \apj, 667, 131, \dodoi{10.1086/520497}

\bibitem[{{Gromadzki} {et~al.}(2019){Gromadzki}, {Hamanowicz}, {Wyrzykowski}, {Sokolovsky}, {Fraser}, {Koz{\l}owski}, {Guillochon}, {Arcavi}, {Trakhtenbrot}, {Jonker}, {Mattila}, {Udalski}, {Szyma{\'n}ski}, {Soszy{\'n}ski}, {Poleski}, {Pietrukowicz}, {Skowron}, {Mr{\'o}z}, {Ulaczyk}, {Pawlak}, {Rybicki}, {Sollerman}, {Taddia}, {Kostrzewa-Rutkowska}, {Onori}, {Young}, {Maguire}, {Smartt}, {Inserra}, {Gal-Yam}, {Rau}, {Chen}, {Angus}, \& {Buckley}}]{Gromadzki2019}
{Gromadzki}, M., {Hamanowicz}, A., {Wyrzykowski}, L., {et~al.} 2019, \aap, 622, L2, \dodoi{10.1051/0004-6361/201833682}

\bibitem[{{Guillochon} {et~al.}(2014){Guillochon}, {Manukian}, \& {Ramirez-Ruiz}}]{Guillochon2014}
{Guillochon}, J., {Manukian}, H., \& {Ramirez-Ruiz}, E. 2014, \apj, 783, 23, \dodoi{10.1088/0004-637X/783/1/23}

\bibitem[{{Guillochon} \& {Ramirez-Ruiz}(2013)}]{Guillochon2013}
{Guillochon}, J., \& {Ramirez-Ruiz}, E. 2013, \apj, 767, 25, \dodoi{10.1088/0004-637X/767/1/25}

\bibitem[{{Guo} {et~al.}(2016){Guo}, {Li}, {Li}, {Daughton}, {Zhang}, {Lloyd-Ronning}, {Liu}, {Zhang}, \& {Deng}}]{Guofan2016}
{Guo}, F., {Li}, X., {Li}, H., {et~al.} 2016, \apjl, 818, L9, \dodoi{10.3847/2041-8205/818/1/L9}

\bibitem[{{Guo} {et~al.}(2023{\natexlab{a}}){Guo}, {Sun}, {Li}, {Jiang}, {Wang}, {Bu}, {Jiang}, {Wang}, {Yao}, {Shen}, {Gu}, \& {Sun}}]{Guohx2023}
{Guo}, H., {Sun}, J., {Li}, S.-L., {et~al.} 2023{\natexlab{a}}, arXiv e-prints, arXiv:2312.06771, \dodoi{10.48550/arXiv.2312.06771}

\bibitem[{{Guo} {et~al.}(2022){Guo}, {Li}, {Zhang}, {Ho}, \& {Wang}}]{Guo2022}
{Guo}, W.-J., {Li}, Y.-R., {Zhang}, Z.-X., {Ho}, L.~C., \& {Wang}, J.-M. 2022, \apj, 929, 19, \dodoi{10.3847/1538-4357/ac4e84}

\bibitem[{{Guo} {et~al.}(2023{\natexlab{b}}){Guo}, {Zou}, {Fawcett}, {Canning}, {Juneau}, {Davis}, {Alexander}, {Jiang}, {Aguilar}, {Ahlen}, {Brooks}, {Claybaugh}, {de la Macorra}, {Doel}, {Fanning}, {Forero-Romero}, {Gontcho}, {Honscheid}, {Kisner}, {Kremin}, {Landriau}, {Meisner}, {Miquel}, {Moustakas}, {Nie}, {Pan}, {Poppett}, {Prada}, {Rezaie}, {Rossi}, {Siudek}, {Sanchez}, {Schubnell}, {Seo}, {Sui}, {Tarl{\'e}}, \& {Zhou}}]{Guo2023}
{Guo}, W.-J., {Zou}, H., {Fawcett}, V.~A., {et~al.} 2023{\natexlab{b}}, arXiv e-prints, arXiv:2307.08289, \dodoi{10.48550/arXiv.2307.08289}

\bibitem[{{Guo} {et~al.}(2024){Guo}, {Zou}, {Greenwell}, {Alexander}, {Fawcett}, {Pan}, {Siudek}, {Aguilar}, {Ahlen}, {Brooks}, {Claybaugh}, {Dawson}, {De La Macorra}, {Doel}, {Font-Ribera}, {Gaztanaga}, {Gontcho}, {Gutierrez}, {Kehoe}, {Kisner}, {Landriau}, {Le Guillou}, {Manera}, {Meisner}, {Mique}, {Moustakas}, {Prada}, {Rossi}, {Sanchez}, {Schubnell}, {Sprayberry}, {Sui}, {Tarle}, {Weaver}, {Xiao}, \& {Zou}}]{GuoWJ2024}
{Guo}, W.-J., {Zou}, H., {Greenwell}, C.~L., {et~al.} 2024, arXiv e-prints, arXiv:2408.00402, \dodoi{10.48550/arXiv.2408.00402}

\bibitem[{{Hammerstein} {et~al.}(2023){Hammerstein}, {van Velzen}, {Gezari}, {Cenko}, {Yao}, {Ward}, {Frederick}, {Villanueva}, {Somalwar}, {Graham}, {Kulkarni}, {Stern}, {Andreoni}, {Bellm}, {Dekany}, {Dhawan}, {Drake}, {Fremling}, {Gatkine}, {Groom}, {Ho}, {Kasliwal}, {Karambelkar}, {Kool}, {Masci}, {Medford}, {Perley}, {Purdum}, {van Roestel}, {Sharma}, {Sollerman}, {Taggart}, \& {Yan}}]{Hammerstein2023}
{Hammerstein}, E., {van Velzen}, S., {Gezari}, S., {et~al.} 2023, \apj, 942, 9, \dodoi{10.3847/1538-4357/aca283}

\bibitem[{{Hinkle} {et~al.}(2024){Hinkle}, {Shappee}, {Auchettl}, {Kochanek}, {Neustadt}, {Polin}, {Strader}, {Holoien}, {Huber}, {Tucker}, {Ashall}, {de Jaeger}, {Desai}, {Do}, {Hoogendam}, \& {Payne}}]{HinkleJ2024}
{Hinkle}, J.~T., {Shappee}, B.~J., {Auchettl}, K., {et~al.} 2024, arXiv e-prints, arXiv:2405.08855, \dodoi{10.48550/arXiv.2405.08855}

\bibitem[{{Holoien} {et~al.}(2022){Holoien}, {Neustadt}, {Vallely}, {Auchettl}, {Hinkle}, {Romero-Ca{\~n}izales}, {Shappee}, {Kochanek}, {Stanek}, {Chen}, {Dong}, {Prieto}, {Thompson}, {Brink}, {Filippenko}, {Zheng}, {Bersier}, {Bose}, {Burgasser}, {Channa}, {de Jaeger}, {Hestenes}, {Im}, {Jeffers}, {Jun}, {Lansbury}, {Post}, {Ross}, {Stern}, {Tang}, {Tucker}, {Valenti}, {Yunus}, \& {Zhang}}]{Holoien2022ApJ}
{Holoien}, T. W.~S., {Neustadt}, J. M.~M., {Vallely}, P.~J., {et~al.} 2022, \apj, 933, 196, \dodoi{10.3847/1538-4357/ac74b9}

\bibitem[{{Hopkins} {et~al.}(2006){Hopkins}, {Hernquist}, {Cox}, {Di Matteo}, {Robertson}, \& {Springel}}]{Hopkins2006}
{Hopkins}, P.~F., {Hernquist}, L., {Cox}, T.~J., {et~al.} 2006, \apjs, 163, 1, \dodoi{10.1086/499298}

\bibitem[{{Hosseinzadeh} {et~al.}(2022){Hosseinzadeh}, {Berger}, {Metzger}, {Gomez}, {Nicholl}, \& {Blanchard}}]{Hosseinzadeh2022}
{Hosseinzadeh}, G., {Berger}, E., {Metzger}, B.~D., {et~al.} 2022, \apj, 933, 14, \dodoi{10.3847/1538-4357/ac67dd}

\bibitem[{{Hovatta} \& {Lindfors}(2019)}]{Hovatta2019}
{Hovatta}, T., \& {Lindfors}, E. 2019, \nar, 87, 101541, \dodoi{10.1016/j.newar.2020.101541}

\bibitem[{{Huang} {et~al.}(2019){Huang}, {Hu}, {Zhao}, {Zhang}, {Lu}, {Wang}, {Zhang}, {Du}, {Li}, {Bai}, {Ho}, {Bian}, {Yuan}, \& {Wang}}]{HuangY2019}
{Huang}, Y.-K., {Hu}, C., {Zhao}, Y.-L., {et~al.} 2019, \apj, 876, 102, \dodoi{10.3847/1538-4357/ab16ef}

\bibitem[{{Hughes} {et~al.}(1985){Hughes}, {Aller}, \& {Aller}}]{Hughes1985}
{Hughes}, P.~A., {Aller}, H.~D., \& {Aller}, M.~F. 1985, \apj, 298, 301, \dodoi{10.1086/163611}

\bibitem[{{Inserra} {et~al.}(2018){Inserra}, {Smartt}, {Gall}, {Leloudas}, {Chen}, {Schulze}, {Jerkstrand}, {Nicholl}, {Anderson}, {Arcavi}, {Benetti}, {Cartier}, {Childress}, {Della Valle}, {Flewelling}, {Fraser}, {Gal-Yam}, {Guti{\'e}rrez}, {Hosseinzadeh}, {Howell}, {Huber}, {Kankare}, {Kr{\"u}hler}, {Magnier}, {Maguire}, {McCully}, {Prajs}, {Primak}, {Scalzo}, {Schmidt}, {Smith}, {Smith}, {Tucker}, {Valenti}, {Wilman}, {Young}, \& {Yuan}}]{Inserra2018}
{Inserra}, C., {Smartt}, S.~J., {Gall}, E.~E.~E., {et~al.} 2018, \mnras, 475, 1046, \dodoi{10.1093/mnras/stx3179}

\bibitem[{{Jiang} {et~al.}(2008){Jiang}, {Fan}, \& {Vestergaard}}]{jianglh2008}
{Jiang}, L., {Fan}, X., \& {Vestergaard}, M. 2008, \apj, 679, 962, \dodoi{10.1086/587868}

\bibitem[{{Jiang} {et~al.}(2019){Jiang}, {Wang}, {Mou}, {Liu}, {Dou}, {Sheng}, \& {Wang}}]{JiangN2019}
{Jiang}, N., {Wang}, T., {Mou}, G., {et~al.} 2019, \apj, 871, 15, \dodoi{10.3847/1538-4357/aaf6b2}

\bibitem[{{Jiang} {et~al.}(2017){Jiang}, {Wang}, {Yan}, {Xiao}, {Yang}, {Dou}, {Wang}, {Cutri}, \& {Mainzer}}]{Jiang2017}
{Jiang}, N., {Wang}, T., {Yan}, L., {et~al.} 2017, \apj, 850, 63, \dodoi{10.3847/1538-4357/aa93f5}

\bibitem[{{Kagan} {et~al.}(2015){Kagan}, {Sironi}, {Cerutti}, \& {Giannios}}]{Kagan2015}
{Kagan}, D., {Sironi}, L., {Cerutti}, B., \& {Giannios}, D. 2015, \ssr, 191, 545, \dodoi{10.1007/s11214-014-0132-9}

\bibitem[{{Kankare} {et~al.}(2017){Kankare}, {Kotak}, {Mattila}, {Lundqvist}, {Ward}, {Fraser}, {Lawrence}, {Smartt}, {Meikle}, {Bruce}, {Harmanen}, {Hutton}, {Inserra}, {Kangas}, {Pastorello}, {Reynolds}, {Romero-Ca{\~n}izales}, {Smith}, {Valenti}, {Chambers}, {Hodapp}, {Huber}, {Kaiser}, {Kudritzki}, {Magnier}, {Tonry}, {Wainscoat}, \& {Waters}}]{Kankare2017}
{Kankare}, E., {Kotak}, R., {Mattila}, S., {et~al.} 2017, Nature Astronomy, 1, 865, \dodoi{10.1038/s41550-017-0290-2}

\bibitem[{{Kara} {et~al.}(2021){Kara}, {Mehdipour}, {Kriss}, {Cackett}, {Arav}, {Barth}, {Byun}, {Brotherton}, {De Rosa}, {Gelbord}, {Hernandez Santisteban}, {Hu}, {Kaastra}, {Landt}, {Li}, {Miller}, {Montano}, {Partington}, {Aceituno}, {Bai}, {Bao}, {Bentz}, {Brink}, {Chelouche}, {Chen}, {Dalla Bonta}, {Dehghanian}, {Du}, {Edelson}, {Ferland}, {Ferrarese}, {Fian}, {Filippenko}, {Fischer}, {Goad}, {Gonzalez Buitrago}, {Gorjian}, {Grier}, {Guo}, {Hall}, {Homayouni}, {Horne}, {Ilic}, {Jiang}, {Joner}, {Kaspi}, {Kochanek}, {Korista}, {Kynoch}, {Li}, {Liu}, {Mc Hardy}, {McLane}, {Mitchell}, {Netzer}, {Olson}, {Pogge}, {Popovic}, {Proga}, {Storchi-Bergmann}, {Strasburger}, {Treu}, {Vestergaard}, {Wang}, {Ward}, {Waters}, {Williams}, {Yang}, {Yao}, {Zastrocky}, {Zhai}, \& {Zu}}]{Kara2021}
{Kara}, E., {Mehdipour}, M., {Kriss}, G.~A., {et~al.} 2021, arXiv e-prints, arXiv:2105.05840.
\newblock \doarXiv{2105.05840}

\bibitem[{{Karas} \& {{\v{S}}ubr}(2007)}]{KarasV2007}
{Karas}, V., \& {{\v{S}}ubr}, L. 2007, \aap, 470, 11, \dodoi{10.1051/0004-6361:20066068}

\bibitem[{{Kaur} \& {Stone}(2024)}]{Kaur2024}
{Kaur}, K., \& {Stone}, N.~C. 2024, arXiv e-prints, arXiv:2405.18500, \dodoi{10.48550/arXiv.2405.18500}

\bibitem[{{Kelly} {et~al.}(2009){Kelly}, {Bechtold}, \& {Siemiginowska}}]{Kelly2009}
{Kelly}, B.~C., {Bechtold}, J., \& {Siemiginowska}, A. 2009, \apj, 698, 895, \dodoi{10.1088/0004-637X/698/1/895}

\bibitem[{{Kochanek}(2016)}]{kochanek2016}
{Kochanek}, C.~S. 2016, \mnras, 458, 127, \dodoi{10.1093/mnras/stw267}

\bibitem[{{Kochanek} {et~al.}(2017){Kochanek}, {Shappee}, {Stanek}, {Holoien}, {Thompson}, {Prieto}, {Dong}, {Shields}, {Will}, {Britt}, {Perzanowski}, \& {Pojma{\'n}ski}}]{Kochanek2017}
{Kochanek}, C.~S., {Shappee}, B.~J., {Stanek}, K.~Z., {et~al.} 2017, \pasp, 129, 104502, \dodoi{10.1088/1538-3873/aa80d9}

\bibitem[{{Komossa} {et~al.}(2006){Komossa}, {Voges}, {Xu}, {Mathur}, {Adorf}, {Lemson}, {Duschl}, \& {Grupe}}]{Komossa2006}
{Komossa}, S., {Voges}, W., {Xu}, D., {et~al.} 2006, \aj, 132, 531, \dodoi{10.1086/505043}

\bibitem[{{Konigl}(1981)}]{Konigl1981}
{Konigl}, A. 1981, \apj, 243, 700, \dodoi{10.1086/158638}

\bibitem[{{Kool} {et~al.}(2020){Kool}, {Reynolds}, {Mattila}, {Kankare}, {P{\'e}rez-Torres}, {Efstathiou}, {Ryder}, {Romero-Ca{\~n}izales}, {Lu}, {Heikkil{\"a}}, {Anderson}, {Berton}, {Bright}, {Cannizzaro}, {Eappachen}, {Fraser}, {Gromadzki}, {Jonker}, {Kuncarayakti}, {Lundqvist}, {Maeda}, {McDermid}, {Medling}, {Moran}, {Reguitti}, {Shahbandeh}, {Tsygankov}, {U}, \& {Wevers}}]{Kool2020}
{Kool}, E.~C., {Reynolds}, T.~M., {Mattila}, S., {et~al.} 2020, \mnras, 498, 2167, \dodoi{10.1093/mnras/staa2351}

\bibitem[{{Koshida} {et~al.}(2014){Koshida}, {Minezaki}, {Yoshii}, {Kobayashi}, {Sakata}, {Sugawara}, {Enya}, {Suganuma}, {Tomita}, {Aoki}, \& {Peterson}}]{Koshida2014}
{Koshida}, S., {Minezaki}, T., {Yoshii}, Y., {et~al.} 2014, \apj, 788, 159, \dodoi{10.1088/0004-637X/788/2/159}

\bibitem[{{Koz{\l}owski} {et~al.}(2010){Koz{\l}owski}, {Kochanek}, {Udalski}, {Wyrzykowski}, {Soszy{\'n}ski}, {Szyma{\'n}ski}, {Kubiak}, {Pietrzy{\'n}ski}, {Szewczyk}, {Ulaczyk}, {Poleski}, \& {OGLE Collaboration}}]{Kozlowski2010}
{Koz{\l}owski}, S., {Kochanek}, C.~S., {Udalski}, A., {et~al.} 2010, \apj, 708, 927, \dodoi{10.1088/0004-637X/708/2/927}

\bibitem[{{Krolik} {et~al.}(1991){Krolik}, {Horne}, {Kallman}, {Malkan}, {Edelson}, \& {Kriss}}]{Krolik1991}
{Krolik}, J.~H., {Horne}, K., {Kallman}, T.~R., {et~al.} 1991, \apj, 371, 541, \dodoi{10.1086/169918}

\bibitem[{{Larionov} {et~al.}(2013){Larionov}, {Jorstad}, {Marscher}, {Morozova}, {Blinov}, {Hagen-Thorn}, {Konstantinova}, {Kopatskaya}, {Larionova}, {Larionova}, \& {Troitsky}}]{Larionov2013}
{Larionov}, V.~M., {Jorstad}, S.~G., {Marscher}, A.~P., {et~al.} 2013, \apj, 768, 40, \dodoi{10.1088/0004-637X/768/1/40}

\bibitem[{{Leloudas} {et~al.}(2019){Leloudas}, {Dai}, {Arcavi}, {Vreeswijk}, {Mockler}, {Roy}, {Malesani}, {Schulze}, {Wevers}, {Fraser}, {Ramirez-Ruiz}, {Auchettl}, {Burke}, {Cannizzaro}, {Charalampopoulos}, {Chen}, {Cikota}, {Della Valle}, {Galbany}, {Gromadzki}, {Heintz}, {Hiramatsu}, {Jonker}, {Kostrzewa-Rutkowska}, {Maguire}, {Mandel}, {Nicholl}, {Onori}, {Roth}, {Smartt}, {Wyrzykowski}, \& {Young}}]{Leloudas2019}
{Leloudas}, G., {Dai}, L., {Arcavi}, I., {et~al.} 2019, \apj, 887, 218, \dodoi{10.3847/1538-4357/ab5792}

\bibitem[{{Li} {et~al.}(2014){Li}, {Wang}, {Hu}, {Du}, \& {Bai}}]{Li2014}
{Li}, Y.-R., {Wang}, J.-M., {Hu}, C., {Du}, P., \& {Bai}, J.-M. 2014, \apjl, 786, L6, \dodoi{10.1088/2041-8205/786/1/L6}

\bibitem[{{Liao} {et~al.}(2019){Liao}, {Dou}, {Jiang}, {Wang}, {Fan}, \& {Wang}}]{Liaonenghui2019}
{Liao}, N.-H., {Dou}, L.-M., {Jiang}, N., {et~al.} 2019, \apjl, 879, L9, \dodoi{10.3847/2041-8213/ab2893}

\bibitem[{{Lin} \& {Shields}(1986)}]{LinDNC1986}
{Lin}, D.~N.~C., \& {Shields}, G.~A. 1986, \apj, 305, 28, \dodoi{10.1086/164225}

\bibitem[{{Liodakis} {et~al.}(2019){Liodakis}, {Romani}, {Filippenko}, {Kocevski}, \& {Zheng}}]{Liodakis2019}
{Liodakis}, I., {Romani}, R.~W., {Filippenko}, A.~V., {Kocevski}, D., \& {Zheng}, W. 2019, \apj, 880, 32, \dodoi{10.3847/1538-4357/ab26b7}

\bibitem[{{Liu} {et~al.}(2018){Liu}, {Dittmann}, {Shen}, \& {Jiang}}]{liuxin2018}
{Liu}, X., {Dittmann}, A., {Shen}, Y., \& {Jiang}, L. 2018, \apj, 859, 8, \dodoi{10.3847/1538-4357/aabb04}

\bibitem[{{Liu} {et~al.}(2023){Liu}, {Malyali}, {Krumpe}, {Homan}, {Goodwin}, {Grotova}, {Kawka}, {Rau}, {Merloni}, {Anderson}, {Miller-Jones}, {Markowitz}, {Ciroi}, {Di Mille}, {Schramm}, {Tang}, {Buckley}, {Gromadzki}, {Jin}, \& {Buchner}}]{Liuz2023}
{Liu}, Z., {Malyali}, A., {Krumpe}, M., {et~al.} 2023, \aap, 669, A75, \dodoi{10.1051/0004-6361/202244805}

\bibitem[{{Loeb} \& {Ulmer}(1997)}]{Loeb1997}
{Loeb}, A., \& {Ulmer}, A. 1997, \apj, 489, 573, \dodoi{10.1086/304814}

\bibitem[{{Lu} {et~al.}(2016){Lu}, {Kumar}, \& {Evans}}]{Luwb2016}
{Lu}, W., {Kumar}, P., \& {Evans}, N.~J. 2016, \mnras, 458, 575, \dodoi{10.1093/mnras/stw307}

\bibitem[{{Lyu} {et~al.}(2019){Lyu}, {Rieke}, \& {Smith}}]{Lyujianwei2019}
{Lyu}, J., {Rieke}, G.~H., \& {Smith}, P.~S. 2019, \apj, 886, 33, \dodoi{10.3847/1538-4357/ab481d}

\bibitem[{{MacLeod} {et~al.}(2010){MacLeod}, {Ivezi{\'c}}, {Kochanek}, {Koz{\l}owski}, {Kelly}, {Bullock}, {Kimball}, {Sesar}, {Westman}, {Brooks}, {Gibson}, {Becker}, \& {de Vries}}]{Macleod2010}
{MacLeod}, C.~L., {Ivezi{\'c}}, {\v{Z}}., {Kochanek}, C.~S., {et~al.} 2010, \apj, 721, 1014, \dodoi{10.1088/0004-637X/721/2/1014}

\bibitem[{{MacLeod} {et~al.}(2019){MacLeod}, {Green}, {Anderson}, {Bruce}, {Eracleous}, {Graham}, {Homan}, {Lawrence}, {LeBleu}, {Ross}, {Ruan}, {Runnoe}, {Stern}, {Burgett}, {Chambers}, {Kaiser}, {Magnier}, \& {Metcalfe}}]{Macleod2019}
{MacLeod}, C.~L., {Green}, P.~J., {Anderson}, S.~F., {et~al.} 2019, \apj, 874, 8, \dodoi{10.3847/1538-4357/ab05e2}

\bibitem[{{MacLeod} {et~al.}(2013){MacLeod}, {Ramirez-Ruiz}, {Grady}, \& {Guillochon}}]{MacleodM2013}
{MacLeod}, M., {Ramirez-Ruiz}, E., {Grady}, S., \& {Guillochon}, J. 2013, \apj, 777, 133, \dodoi{10.1088/0004-637X/777/2/133}

\bibitem[{{Mainetti} {et~al.}(2017){Mainetti}, {Lupi}, {Campana}, {Colpi}, {Coughlin}, {Guillochon}, \& {Ramirez-Ruiz}}]{Mainetti2017}
{Mainetti}, D., {Lupi}, A., {Campana}, S., {et~al.} 2017, \aap, 600, A124, \dodoi{10.1051/0004-6361/201630092}

\bibitem[{{Makrygianni} {et~al.}(2023){Makrygianni}, {Trakhtenbrot}, {Arcavi}, {Ricci}, {Lam}, {Horesh}, {Sfaradi}, {Bostroem}, {Hosseinzadeh}, {Howell}, {Pellegrino}, {Fender}, {Green}, {Williams}, \& {Bright}}]{Makrygianni2023}
{Makrygianni}, L., {Trakhtenbrot}, B., {Arcavi}, I., {et~al.} 2023, \apj, 953, 32, \dodoi{10.3847/1538-4357/ace1ee}

\bibitem[{{Malyali} {et~al.}(2021){Malyali}, {Rau}, {Merloni}, {Nandra}, {Buchner}, {Liu}, {Gezari}, {Sollerman}, {Shappee}, {Trakhtenbrot}, {Arcavi}, {Ricci}, {van Velzen}, {Goobar}, {Frederick}, {Kawka}, {Tartaglia}, {Burke}, {Hiramatsu}, {Schramm}, {van der Boom}, {Anderson}, {Miller-Jones}, {Bellm}, {Drake}, {Duev}, {Fremling}, {Graham}, {Masci}, {Rusholme}, {Soumagnac}, \& {Walters}}]{Malyali2021}
{Malyali}, A., {Rau}, A., {Merloni}, A., {et~al.} 2021, \aap, 647, A9, \dodoi{10.1051/0004-6361/202039681}

\bibitem[{{Malyali} {et~al.}(2023){Malyali}, {Liu}, {Rau}, {Grotova}, {Merloni}, {Goodwin}, {Anderson}, {Miller-Jones}, {Kawka}, {Arcodia}, {Buchner}, {Nandra}, {Homan}, \& {Krumpe}}]{Malyali2023}
{Malyali}, A., {Liu}, Z., {Rau}, A., {et~al.} 2023, \mnras, 520, 3549, \dodoi{10.1093/mnras/stad022}

\bibitem[{{Marscher} \& {Gear}(1985)}]{Marscher1985}
{Marscher}, A.~P., \& {Gear}, W.~K. 1985, \apj, 298, 114, \dodoi{10.1086/163592}

\bibitem[{{Masci} {et~al.}(2019){Masci}, {Laher}, {Rusholme}, {Shupe}, {Groom}, {Surace}, {Jackson}, {Monkewitz}, {Beck}, {Flynn}, {Terek}, {Landry}, {Hacopians}, {Desai}, {Howell}, {Brooke}, {Imel}, {Wachter}, {Ye}, {Lin}, {Cenko}, {Cunningham}, {Rebbapragada}, {Bue}, {Miller}, {Mahabal}, {Bellm}, {Patterson}, {Juri{\'c}}, {Golkhou}, {Ofek}, {Walters}, {Graham}, {Kasliwal}, {Dekany}, {Kupfer}, {Burdge}, {Cannella}, {Barlow}, {Van Sistine}, {Giomi}, {Fremling}, {Blagorodnova}, {Levitan}, {Riddle}, {Smith}, {Helou}, {Prince}, \& {Kulkarni}}]{Masci2019}
{Masci}, F.~J., {Laher}, R.~R., {Rusholme}, B., {et~al.} 2019, \pasp, 131, 018003, \dodoi{10.1088/1538-3873/aae8ac}

\bibitem[{{Mattila} {et~al.}(2018){Mattila}, {P{\'e}rez-Torres}, {Efstathiou}, {Mimica}, {Fraser}, {Kankare}, {Alberdi}, {Aloy}, {Heikkil{\"a}}, {Jonker}, {Lundqvist}, {Mart{\'\i}-Vidal}, {Meikle}, {Romero-Ca{\~n}izales}, {Smartt}, {Tsygankov}, {Varenius}, {Alonso-Herrero}, {Bondi}, {Fransson}, {Herrero-Illana}, {Kangas}, {Kotak}, {Ram{\'\i}rez-Olivencia}, {V{\"a}is{\"a}nen}, {Beswick}, {Clements}, {Greimel}, {Harmanen}, {Kotilainen}, {Nandra}, {Reynolds}, {Ryder}, {Walton}, {Wiik}, \& {{\"O}stlin}}]{Mattila2018}
{Mattila}, S., {P{\'e}rez-Torres}, M., {Efstathiou}, A., {et~al.} 2018, Science, 361, 482, \dodoi{10.1126/science.aao4669}

\bibitem[{{McKernan} {et~al.}(2022){McKernan}, {Ford}, {Cantiello}, {Graham}, {Jermyn}, {Leigh}, {Ryu}, \& {Stern}}]{MckernanB2022}
{McKernan}, B., {Ford}, K.~E.~S., {Cantiello}, M., {et~al.} 2022, \mnras, 514, 4102, \dodoi{10.1093/mnras/stac1310}

\bibitem[{{Merloni} {et~al.}(2015){Merloni}, {Dwelly}, {Salvato}, {Georgakakis}, {Greiner}, {Krumpe}, {Nandra}, {Ponti}, \& {Rau}}]{Merloni2015}
{Merloni}, A., {Dwelly}, T., {Salvato}, M., {et~al.} 2015, \mnras, 452, 69, \dodoi{10.1093/mnras/stv1095}

\bibitem[{{Metzger}(2022)}]{Metzger2022}
{Metzger}, B.~D. 2022, \apjl, 937, L12, \dodoi{10.3847/2041-8213/ac90ba}

\bibitem[{{Metzger} \& {Stone}(2016)}]{Metzger2016}
{Metzger}, B.~D., \& {Stone}, N.~C. 2016, \mnras, 461, 948, \dodoi{10.1093/mnras/stw1394}

\bibitem[{{Miles} {et~al.}(2020){Miles}, {Coughlin}, \& {Nixon}}]{Miles2020}
{Miles}, P.~R., {Coughlin}, E.~R., \& {Nixon}, C.~J. 2020, \apj, 899, 36, \dodoi{10.3847/1538-4357/ab9c9f}

\bibitem[{{Miller} {et~al.}(2009){Miller}, {Chornock}, {Perley}, {Ganeshalingam}, {Li}, {Butler}, {Bloom}, {Smith}, {Modjaz}, {Poznanski}, {Filippenko}, {Griffith}, {Shiode}, \& {Silverman}}]{Miller2009}
{Miller}, A.~A., {Chornock}, R., {Perley}, D.~A., {et~al.} 2009, \apj, 690, 1303, \dodoi{10.1088/0004-637X/690/2/1303}

\bibitem[{{Miller}(2015)}]{Miller2015}
{Miller}, M.~C. 2015, \apj, 805, 83, \dodoi{10.1088/0004-637X/805/1/83}

\bibitem[{{Minezaki} {et~al.}(2019){Minezaki}, {Yoshii}, {Kobayashi}, {Sugawara}, {Sakata}, {Enya}, {Koshida}, {Tomita}, {Suganuma}, {Aoki}, \& {Peterson}}]{minezaki2019}
{Minezaki}, T., {Yoshii}, Y., {Kobayashi}, Y., {et~al.} 2019, \apj, 886, 150, \dodoi{10.3847/1538-4357/ab4f7b}

\bibitem[{{Mockler} {et~al.}(2024){Mockler}, {Gallegos-Garcia}, {G{\"o}tberg}, {Miller}, \& {Ramirez-Ruiz}}]{mockler2024}
{Mockler}, B., {Gallegos-Garcia}, M., {G{\"o}tberg}, Y., {Miller}, J., \& {Ramirez-Ruiz}, E. 2024, arXiv e-prints, arXiv:2406.04455, \dodoi{10.48550/arXiv.2406.04455}

\bibitem[{{Netzer}(2013)}]{NetzerH2013}
{Netzer}, H. 2013, {The Physics and Evolution of Active Galactic Nuclei}

\bibitem[{{Netzer} {et~al.}(1985){Netzer}, {Elitzur}, \& {Ferland}}]{Netzer1985}
{Netzer}, H., {Elitzur}, M., \& {Ferland}, G.~J. 1985, \apj, 299, 752, \dodoi{10.1086/163741}

\bibitem[{{Neugebauer} \& {Matthews}(1999)}]{Neugebauer1999}
{Neugebauer}, G., \& {Matthews}, K. 1999, \aj, 118, 35, \dodoi{10.1086/300945}

\bibitem[{{Neustadt} {et~al.}(2020){Neustadt}, {Holoien}, {Kochanek}, {Auchettl}, {Brown}, {Shappee}, {Pogge}, {Dong}, {Stanek}, {Tucker}, {Bose}, {Chen}, {Ricci}, {Vallely}, {Prieto}, {Thompson}, {Coulter}, {Drout}, {Foley}, {Kilpatrick}, {Piro}, {Rojas-Bravo}, {Buckley}, {Gromadzki}, {Dimitriadis}, {Siebert}, {Do}, {Huber}, \& {Payne}}]{Neustadt2020}
{Neustadt}, J.~M.~M., {Holoien}, T.~W.~S., {Kochanek}, C.~S., {et~al.} 2020, \mnras, 494, 2538, \dodoi{10.1093/mnras/staa859}

\bibitem[{{Nicholl} {et~al.}(2019){Nicholl}, {Short}, {Lawrence}, {Ross}, \& {Smartt}}]{NichollM2019}
{Nicholl}, M., {Short}, P., {Lawrence}, A., {Ross}, N., \& {Smartt}, S. 2019, Transient Name Server Classification Report, 2019-1586, 1

\bibitem[{{Osmer}(1980)}]{Osmer1980}
{Osmer}, P.~S. 1980, \apj, 237, 666, \dodoi{10.1086/157913}

\bibitem[{{Osterbrock}(1974)}]{osterbrock1974}
{Osterbrock}, D.~E. 1974, {Astrophysics of gaseous nebulae}

\bibitem[{{Paltani} {et~al.}(1997){Paltani}, {Courvoisier}, {Blecha}, \& {Bratschi}}]{Paltani1997}
{Paltani}, S., {Courvoisier}, T.~J.~L., {Blecha}, A., \& {Bratschi}, P. 1997, \aap, 327, 539, \dodoi{10.48550/arXiv.astro-ph/9706203}

\bibitem[{{Payne} {et~al.}(2021){Payne}, {Shappee}, {Hinkle}, {Vallely}, {Kochanek}, {Holoien}, {Auchettl}, {Stanek}, {Thompson}, {Neustadt}, {Tucker}, {Armstrong}, {Brimacombe}, {Cacella}, {Cornect}, {Denneau}, {Fausnaugh}, {Flewelling}, {Grupe}, {Heinze}, {Lopez}, {Monard}, {Prieto}, {Schneider}, {Sheppard}, {Tonry}, \& {Weiland}}]{Payne2021}
{Payne}, A.~V., {Shappee}, B.~J., {Hinkle}, J.~T., {et~al.} 2021, \apj, 910, 125, \dodoi{10.3847/1538-4357/abe38d}

\bibitem[{{Perger} {et~al.}(2023){Perger}, {Frey}, \& {Gab{\'a}nyi}}]{Perger2023}
{Perger}, K., {Frey}, S., \& {Gab{\'a}nyi}, K.~{\'E}. 2023, \apss, 368, 18, \dodoi{10.1007/s10509-023-04176-4}

\bibitem[{{Peterson}(1993)}]{Peterson1993}
{Peterson}, B.~M. 1993, \pasp, 105, 247, \dodoi{10.1086/133140}

\bibitem[{{Peterson} {et~al.}(1998){Peterson}, {Wanders}, {Horne}, {Collier}, {Alexander}, {Kaspi}, \& {Maoz}}]{Peterson1998}
{Peterson}, B.~M., {Wanders}, I., {Horne}, K., {et~al.} 1998, \pasp, 110, 660, \dodoi{10.1086/316177}

\bibitem[{{Petrushevska} {et~al.}(2023){Petrushevska}, {Leloudas}, {Ili{\'c}}, {Bronikowski}, {Charalampopoulos}, {Jaisawal}, {Paraskeva}, {Pursiainen}, {Raki{\'c}}, {Schulze}, {Taggart}, {Wedderkopp}, {Anderson}, {de Boer}, {Chambers}, {Chen}, {Damljanovi{\'c}}, {Fraser}, {Gao}, {Gomboc}, {Gromadzki}, {Ihanec}, {Maguire}, {Mar{\v{c}}un}, {M{\"u}ller-Bravo}, {Nicholl}, {Onori}, {Reynolds}, {Smartt}, {Sollerman}, {Smith}, {Wevers}, \& {Wyrzykowski}}]{Petrushevska2023}
{Petrushevska}, T., {Leloudas}, G., {Ili{\'c}}, D., {et~al.} 2023, \aap, 669, A140, \dodoi{10.1051/0004-6361/202244623}

\bibitem[{{Phinney}(1989)}]{Phinney1989}
{Phinney}, E.~S. 1989, in The Center of the Galaxy, ed. M.~{Morris}, Vol. 136, 543

\bibitem[{{Piran} {et~al.}(2015){Piran}, {Svirski}, {Krolik}, {Cheng}, \& {Shiokawa}}]{Piran2015}
{Piran}, T., {Svirski}, G., {Krolik}, J., {Cheng}, R.~M., \& {Shiokawa}, H. 2015, \apj, 806, 164, \dodoi{10.1088/0004-637X/806/2/164}

\bibitem[{{Planck Collaboration} {et~al.}(2020){Planck Collaboration}, {Aghanim}, {Akrami}, {Ashdown}, {Aumont}, {Baccigalupi}, {Ballardini}, {Banday}, {Barreiro}, {Bartolo}, {Basak}, {Battye}, {Benabed}, {Bernard}, {Bersanelli}, {Bielewicz}, {Bock}, {Bond}, {Borrill}, {Bouchet}, {Boulanger}, {Bucher}, {Burigana}, {Butler}, {Calabrese}, {Cardoso}, {Carron}, {Challinor}, {Chiang}, {Chluba}, {Colombo}, {Combet}, {Contreras}, {Crill}, {Cuttaia}, {de Bernardis}, {de Zotti}, {Delabrouille}, {Delouis}, {Di Valentino}, {Diego}, {Dor{\'e}}, {Douspis}, {Ducout}, {Dupac}, {Dusini}, {Efstathiou}, {Elsner}, {En{\ss}lin}, {Eriksen}, {Fantaye}, {Farhang}, {Fergusson}, {Fernandez-Cobos}, {Finelli}, {Forastieri}, {Frailis}, {Fraisse}, {Franceschi}, {Frolov}, {Galeotta}, {Galli}, {Ganga}, {G{\'e}nova-Santos}, {Gerbino}, {Ghosh}, {Gonz{\'a}lez-Nuevo}, {G{\'o}rski}, {Gratton}, {Gruppuso}, {Gudmundsson}, {Hamann}, {Handley}, {Hansen}, {Herranz}, {Hildebrandt}, {Hivon}, {Huang}, {Jaffe}, {Jones}, {Karakci}, {Keih{\"a}nen},
  {Keskitalo}, {Kiiveri}, {Kim}, {Kisner}, {Knox}, {Krachmalnicoff}, {Kunz}, {Kurki-Suonio}, {Lagache}, {Lamarre}, {Lasenby}, {Lattanzi}, {Lawrence}, {Le Jeune}, {Lemos}, {Lesgourgues}, {Levrier}, {Lewis}, {Liguori}, {Lilje}, {Lilley}, {Lindholm}, {L{\'o}pez-Caniego}, {Lubin}, {Ma}, {Mac{\'\i}as-P{\'e}rez}, {Maggio}, {Maino}, {Mandolesi}, {Mangilli}, {Marcos-Caballero}, {Maris}, {Martin}, {Martinelli}, {Mart{\'\i}nez-Gonz{\'a}lez}, {Matarrese}, {Mauri}, {McEwen}, {Meinhold}, {Melchiorri}, {Mennella}, {Migliaccio}, {Millea}, {Mitra}, {Miville-Desch{\^e}nes}, {Molinari}, {Montier}, {Morgante}, {Moss}, {Natoli}, {N{\o}rgaard-Nielsen}, {Pagano}, {Paoletti}, {Partridge}, {Patanchon}, {Peiris}, {Perrotta}, {Pettorino}, {Piacentini}, {Polastri}, {Polenta}, {Puget}, {Rachen}, {Reinecke}, {Remazeilles}, {Renzi}, {Rocha}, {Rosset}, {Roudier}, {Rubi{\~n}o-Mart{\'\i}n}, {Ruiz-Granados}, {Salvati}, {Sandri}, {Savelainen}, {Scott}, {Shellard}, {Sirignano}, {Sirri}, {Spencer}, {Sunyaev}, {Suur-Uski}, {Tauber}, {Tavagnacco},
  {Tenti}, {Toffolatti}, {Tomasi}, {Trombetti}, {Valenziano}, {Valiviita}, {Van Tent}, {Vibert}, {Vielva}, {Villa}, {Vittorio}, {Wandelt}, {Wehus}, {White}, {White}, {Zacchei}, \& {Zonca}}]{Planck2020}
{Planck Collaboration}, {Aghanim}, N., {Akrami}, Y., {et~al.} 2020, \aap, 641, A6, \dodoi{10.1051/0004-6361/201833910}

\bibitem[{{Prasad} {et~al.}(2024){Prasad}, {Wang}, {Perna}, {Ford}, \& {McKernan}}]{Prasad2024}
{Prasad}, C., {Wang}, Y., {Perna}, R., {Ford}, K.~E.~S., \& {McKernan}, B. 2024, \mnras, 531, 1409, \dodoi{10.1093/mnras/stae1263}

\bibitem[{{Raiteri} {et~al.}(2003){Raiteri}, {Villata}, {Tosti}, {Nesci}, {Massaro}, {Aller}, {Aller}, {Ter{\"a}sranta}, {Kurtanidze}, {Nikolashvili}, {Ibrahimov}, {Papadakis}, {Krichbaum}, {Kraus}, {Witzel}, {Ungerechts}, {Lisenfeld}, {Bach}, {Cim{\`o}}, {Ciprini}, {Fuhrmann}, {Kimeridze}, {Lanteri}, {Maesano}, {Montagni}, {Nucciarelli}, \& {Ostorero}}]{Raiteri2003}
{Raiteri}, C.~M., {Villata}, M., {Tosti}, G., {et~al.} 2003, \aap, 402, 151, \dodoi{10.1051/0004-6361:20030256}

\bibitem[{{Raiteri} {et~al.}(2006){Raiteri}, {Villata}, {Kadler}, {Ibrahimov}, {Kurtanidze}, {Larionov}, {Tornikoski}, {Boltwood}, {Lee}, {Aller}, {Romero}, {Aller}, {Araudo}, {Arkharov}, {Bach}, {Barnaby}, {Berdyugin}, {Buemi}, {Carini}, {Carosati}, {Cellone}, {Cool}, {Dolci}, {Efimova}, {Fuhrmann}, {Hagen-Thorn}, {Holcomb}, {Ilyin}, {Impellizzeri}, {Ivanidze}, {Kapanadze}, {Kerp}, {Konstantinova}, {Kovalev}, {Kovalev}, {Kraus}, {Krichbaum}, {L{\"a}hteenm{\"a}ki}, {Lanteri}, {Leto}, {Lindfors}, {Mattox}, {Napoleone}, {Nikolashvili}, {Nilsson}, {Ohlert}, {Papadakis}, {Pasanen}, {Poteet}, {Pursimo}, {Ros}, {Sigua}, {Smith}, {Takalo}, {Trigilio}, {Tr{\"o}ller}, {Umana}, {Ungerechts}, {Walters}, {Witzel}, \& {Xilouris}}]{Raiteri2006}
{Raiteri}, C.~M., {Villata}, M., {Kadler}, M., {et~al.} 2006, \aap, 459, 731, \dodoi{10.1051/0004-6361:20065744}

\bibitem[{{Rakshit} {et~al.}(2017){Rakshit}, {Stalin}, {Chand}, \& {Zhang}}]{Rakshit2017}
{Rakshit}, S., {Stalin}, C.~S., {Chand}, H., \& {Zhang}, X.-G. 2017, \apjs, 229, 39, \dodoi{10.3847/1538-4365/aa6971}

\bibitem[{{Rani} {et~al.}(2010){Rani}, {Gupta}, {Joshi}, {Ganesh}, \& {Wiita}}]{Rani2010}
{Rani}, B., {Gupta}, A.~C., {Joshi}, U.~C., {Ganesh}, S., \& {Wiita}, P.~J. 2010, \apjl, 719, L153, \dodoi{10.1088/2041-8205/719/2/L153}

\bibitem[{{Rani} {et~al.}(2013){Rani}, {Krichbaum}, {Fuhrmann}, {B{\"o}ttcher}, {Lott}, {Aller}, {Aller}, {Angelakis}, {Bach}, {Bastieri}, {Falcone}, {Fukazawa}, {Gabanyi}, {Gupta}, {Gurwell}, {Itoh}, {Kawabata}, {Krips}, {L{\"a}hteenm{\"a}ki}, {Liu}, {Marchili}, {Max-Moerbeck}, {Nestoras}, {Nieppola}, {Quintana-Lacaci}, {Readhead}, {Richards}, {Sasada}, {Sievers}, {Sokolovsky}, {Stroh}, {Tammi}, {Tornikoski}, {Uemura}, {Ungerechts}, {Urano}, \& {Zensus}}]{raniB2013}
{Rani}, B., {Krichbaum}, T.~P., {Fuhrmann}, L., {et~al.} 2013, \aap, 552, A11, \dodoi{10.1051/0004-6361/201321058}

\bibitem[{{Rees}(1988)}]{Rees1988}
{Rees}, M.~J. 1988, \nat, 333, 523, \dodoi{10.1038/333523a0}

\bibitem[{{Reynolds} {et~al.}(2022){Reynolds}, {Mattila}, {Efstathiou}, {Kankare}, {Kool}, {Ryder}, {Pe{\~n}a-Mo{\~n}ino}, \& {P{\'e}rez-Torres}}]{Reynolds2022}
{Reynolds}, T.~M., {Mattila}, S., {Efstathiou}, A., {et~al.} 2022, \aap, 664, A158, \dodoi{10.1051/0004-6361/202243289}

\bibitem[{{Ricci} \& {Trakhtenbrot}(2023)}]{Ricci2023}
{Ricci}, C., \& {Trakhtenbrot}, B. 2023, Nature Astronomy, 7, 1282, \dodoi{10.1038/s41550-023-02108-4}

\bibitem[{{Roth} \& {Kasen}(2018)}]{RothN2018}
{Roth}, N., \& {Kasen}, D. 2018, \apj, 855, 54, \dodoi{10.3847/1538-4357/aaaec6}

\bibitem[{{Roth} {et~al.}(2016){Roth}, {Kasen}, {Guillochon}, \& {Ramirez-Ruiz}}]{Roth2016}
{Roth}, N., {Kasen}, D., {Guillochon}, J., \& {Ramirez-Ruiz}, E. 2016, \apj, 827, 3, \dodoi{10.3847/0004-637X/827/1/3}

\bibitem[{{Sanders} {et~al.}(1988){Sanders}, {Soifer}, {Elias}, {Neugebauer}, \& {Matthews}}]{Sanders1988}
{Sanders}, D.~B., {Soifer}, B.~T., {Elias}, J.~H., {Neugebauer}, G., \& {Matthews}, K. 1988, \apjl, 328, L35, \dodoi{10.1086/185155}

\bibitem[{{Shiokawa} {et~al.}(2015){Shiokawa}, {Krolik}, {Cheng}, {Piran}, \& {Noble}}]{shiokawa2015}
{Shiokawa}, H., {Krolik}, J.~H., {Cheng}, R.~M., {Piran}, T., \& {Noble}, S.~C. 2015, \apj, 804, 85, \dodoi{10.1088/0004-637X/804/2/85}

\bibitem[{{Singh} \& {Chand}(2018)}]{Singh2018}
{Singh}, V., \& {Chand}, H. 2018, \mnras, 480, 1796, \dodoi{10.1093/mnras/sty1818}

\bibitem[{{Somalwar} {et~al.}(2023){Somalwar}, {Ravi}, {Yao}, {Guolo}, {Graham}, {Hammerstein}, {Lu}, {Nicholl}, {Sharma}, {Stein}, {van Velzen}, {Bellm}, {Coughlin}, {Groom}, {Masci}, \& {Riddle}}]{Somalwar2023}
{Somalwar}, J.~J., {Ravi}, V., {Yao}, Y., {et~al.} 2023, arXiv e-prints, arXiv:2310.03782, \dodoi{10.48550/arXiv.2310.03782}

\bibitem[{{Soraisam} {et~al.}(2022){Soraisam}, {Matheson}, {Lee}, {Saha}, {Narayan}, {Wolf}, {Scott}, {Figuereo}, {Nu{\~n}ez}, {McKinnon}, {Guhathakurta}, {Brink}, {Filippenko}, \& {Smith}}]{SoraisamM2022}
{Soraisam}, M., {Matheson}, T., {Lee}, C.-H., {et~al.} 2022, \apjl, 926, L11, \dodoi{10.3847/2041-8213/ac4e99}

\bibitem[{{Stern} {et~al.}(2005){Stern}, {Eisenhardt}, {Gorjian}, {Kochanek}, {Caldwell}, {Eisenstein}, {Brodwin}, {Brown}, {Cool}, {Dey}, {Green}, {Jannuzi}, {Murray}, {Pahre}, \& {Willner}}]{2005ApJ...631..163S}
{Stern}, D., {Eisenhardt}, P., {Gorjian}, V., {et~al.} 2005, \apj, 631, 163, \dodoi{10.1086/432523}

\bibitem[{{Stern} {et~al.}(2012){Stern}, {Assef}, {Benford}, {Blain}, {Cutri}, {Dey}, {Eisenhardt}, {Griffith}, {Jarrett}, {Lake}, {Masci}, {Petty}, {Stanford}, {Tsai}, {Wright}, {Yan}, {Harrison}, \& {Madsen}}]{SternD2012}
{Stern}, D., {Assef}, R.~J., {Benford}, D.~J., {et~al.} 2012, \apj, 753, 30, \dodoi{10.1088/0004-637X/753/1/30}

\bibitem[{{Stone} {et~al.}(2013){Stone}, {Sari}, \& {Loeb}}]{StoneN2013}
{Stone}, N., {Sari}, R., \& {Loeb}, A. 2013, \mnras, 435, 1809, \dodoi{10.1093/mnras/stt1270}

\bibitem[{{Stone} \& {van Velzen}(2016)}]{Stone2016}
{Stone}, N.~C., \& {van Velzen}, S. 2016, \apjl, 825, L14, \dodoi{10.3847/2041-8205/825/1/L14}

\bibitem[{{Sun} {et~al.}(2024){Sun}, {Jiang}, {Dou}, {Shu}, {Zhu}, {Dong}, {Buckley}, {Cenko}, {Fan}, {Gromadzki}, {Liu}, {Wang}, {Wang}, {Wang}, {Wu}, {Yang}, {Zhang}, {Zhang}, \& {Zhang}}]{SunLM2024}
{Sun}, L., {Jiang}, N., {Dou}, L., {et~al.} 2024, arXiv e-prints, arXiv:2410.09720, \dodoi{10.48550/arXiv.2410.09720}

\bibitem[{{Tadhunter} {et~al.}(2017){Tadhunter}, {Spence}, {Rose}, {Mullaney}, \& {Crowther}}]{Tadhunter2017}
{Tadhunter}, C., {Spence}, R., {Rose}, M., {Mullaney}, J., \& {Crowther}, P. 2017, Nature Astronomy, 1, 0061, \dodoi{10.1038/s41550-017-0061}

\bibitem[{{Terreran} {et~al.}(2016){Terreran}, {Berton}, {Benetti}, {Cappellaro}, {Elias-Rosa}, {Ochner}, {Pastorello}, {Tomasella}, \& {Turatto}}]{Terreran2016}
{Terreran}, G., {Berton}, M., {Benetti}, S., {et~al.} 2016, The Astronomer's Telegram, 9417, 1

\bibitem[{{Tonry} {et~al.}(2018){Tonry}, {Denneau}, {Heinze}, {Stalder}, {Smith}, {Smartt}, {Stubbs}, {Weiland}, \& {Rest}}]{Tonry2018}
{Tonry}, J.~L., {Denneau}, L., {Heinze}, A.~N., {et~al.} 2018, \pasp, 130, 064505, \dodoi{10.1088/1538-3873/aabadf}

\bibitem[{{Trakhtenbrot} {et~al.}(2019){Trakhtenbrot}, {Arcavi}, {Ricci}, {Tacchella}, {Stern}, {Netzer}, {Jonker}, {Horesh}, {Mej{\'\i}a-Restrepo}, {Hosseinzadeh}, {Hallefors}, {Howell}, {McCully}, {Balokovi{\'c}}, {Heida}, {Kamraj}, {Lansbury}, {Wyrzykowski}, {Gromadzki}, {Hamanowicz}, {Cenko}, {Sand}, {Hsiao}, {Phillips}, {Diamond}, {Kara}, {Gendreau}, {Arzoumanian}, \& {Remillard}}]{Trakhtenbrot2019}
{Trakhtenbrot}, B., {Arcavi}, I., {Ricci}, C., {et~al.} 2019, Nature Astronomy, 3, 242, \dodoi{10.1038/s41550-018-0661-3}

\bibitem[{{Urry} \& {Mushotzky}(1982)}]{Urry1982}
{Urry}, C.~M., \& {Mushotzky}, R.~F. 1982, \apj, 253, 38, \dodoi{10.1086/159607}

\bibitem[{{van Velzen} {et~al.}(2020){van Velzen}, {Holoien}, {Onori}, {Hung}, \& {Arcavi}}]{vanvelzen2020}
{van Velzen}, S., {Holoien}, T. W.~S., {Onori}, F., {Hung}, T., \& {Arcavi}, I. 2020, \ssr, 216, 124, \dodoi{10.1007/s11214-020-00753-z}

\bibitem[{{van Velzen} {et~al.}(2011){van Velzen}, {Farrar}, {Gezari}, {Morrell}, {Zaritsky}, {{\"O}stman}, {Smith}, {Gelfand}, \& {Drake}}]{vanvelzen2011}
{van Velzen}, S., {Farrar}, G.~R., {Gezari}, S., {et~al.} 2011, \apj, 741, 73, \dodoi{10.1088/0004-637X/741/2/73}

\bibitem[{{van Velzen} {et~al.}(2019){van Velzen}, {Gezari}, {Cenko}, {Kara}, {Miller-Jones}, {Hung}, {Bright}, {Roth}, {Blagorodnova}, {Huppenkothen}, {Yan}, {Ofek}, {Sollerman}, {Frederick}, {Ward}, {Graham}, {Fender}, {Kasliwal}, {Canella}, {Stein}, {Giomi}, {Brinnel}, {van Santen}, {Nordin}, {Bellm}, {Dekany}, {Fremling}, {Golkhou}, {Kupfer}, {Kulkarni}, {Laher}, {Mahabal}, {Masci}, {Miller}, {Neill}, {Riddle}, {Rigault}, {Rusholme}, {Soumagnac}, \& {Tachibana}}]{vanvelzen2019}
{van Velzen}, S., {Gezari}, S., {Cenko}, S.~B., {et~al.} 2019, \apj, 872, 198, \dodoi{10.3847/1538-4357/aafe0c}

\bibitem[{{van Velzen} {et~al.}(2021){van Velzen}, {Gezari}, {Hammerstein}, {Roth}, {Frederick}, {Ward}, {Hung}, {Cenko}, {Stein}, {Perley}, {Taggart}, {Foley}, {Sollerman}, {Blagorodnova}, {Andreoni}, {Bellm}, {Brinnel}, {De}, {Dekany}, {Feeney}, {Fremling}, {Giomi}, {Golkhou}, {Graham}, {Ho}, {Kasliwal}, {Kilpatrick}, {Kulkarni}, {Kupfer}, {Laher}, {Mahabal}, {Masci}, {Miller}, {Nordin}, {Riddle}, {Rusholme}, {van Santen}, {Sharma}, {Shupe}, \& {Soumagnac}}]{van_ZTF_TDEsample}
{van Velzen}, S., {Gezari}, S., {Hammerstein}, E., {et~al.} 2021, \apj, 908, 4, \dodoi{10.3847/1538-4357/abc258}

\bibitem[{{Vanden Berk} {et~al.}(2001){Vanden Berk}, {Richards}, {Bauer}, {Strauss}, {Schneider}, {Heckman}, {York}, {Hall}, {Fan}, {Knapp}, {Anderson}, {Annis}, {Bahcall}, {Bernardi}, {Briggs}, {Brinkmann}, {Brunner}, {Burles}, {Carey}, {Castander}, {Connolly}, {Crocker}, {Csabai}, {Doi}, {Finkbeiner}, {Friedman}, {Frieman}, {Fukugita}, {Gunn}, {Hennessy}, {Ivezi{\'c}}, {Kent}, {Kunszt}, {Lamb}, {Leger}, {Long}, {Loveday}, {Lupton}, {Meiksin}, {Merelli}, {Munn}, {Newberg}, {Newcomb}, {Nichol}, {Owen}, {Pier}, {Pope}, {Rockosi}, {Schlegel}, {Siegmund}, {Smee}, {Snir}, {Stoughton}, {Stubbs}, {SubbaRao}, {Szalay}, {Szokoly}, {Tremonti}, {Uomoto}, {Waddell}, {Yanny}, \& {Zheng}}]{vandenberk2001}
{Vanden Berk}, D.~E., {Richards}, G.~T., {Bauer}, A., {et~al.} 2001, \aj, 122, 549, \dodoi{10.1086/321167}

\bibitem[{{Vaona} {et~al.}(2012){Vaona}, {Ciroi}, {Di Mille}, {Cracco}, {La Mura}, \& {Rafanelli}}]{Vaona2012}
{Vaona}, L., {Ciroi}, S., {Di Mille}, F., {et~al.} 2012, \mnras, 427, 1266, \dodoi{10.1111/j.1365-2966.2012.22060.x}

\bibitem[{{Veilleux} \& {Osterbrock}(1987)}]{Veilleux1987}
{Veilleux}, S., \& {Osterbrock}, D.~E. 1987, \apjs, 63, 295, \dodoi{10.1086/191166}

\bibitem[{Veres {et~al.}(2024)Veres, Franckowiak, van Velzen, Adebahr, Taziaux, Necker, Stein, Kier, Mueller, Bomans, Jordana-Mitjans, Kowalski, Hammerstein, Marci-Boehncke, Reusch, Garrappa, Rose, \& Das}]{veres2024}
Veres, P.~M., Franckowiak, A., van Velzen, S., {et~al.} 2024, Back from the dead: AT2019aalc as a candidate repeating TDE in an AGN.
\newblock \doarXiv{2408.17419}

\bibitem[{{Wang} {et~al.}(2011){Wang}, {Ge}, {Hu}, {Baldwin}, {Li}, {Ferland}, {Xiang}, {Yan}, \& {Zhang}}]{wangjm2011}
{Wang}, J.-M., {Ge}, J.-Q., {Hu}, C., {et~al.} 2011, \apj, 739, 3, \dodoi{10.1088/0004-637X/739/1/3}

\bibitem[{{Wang} {et~al.}(2014){Wang}, {Du}, {Hu}, {Netzer}, {Bai}, {Lu}, {Kaspi}, {Qiu}, {Li}, {Wang}, \& {SEAMBH Collaboration}}]{WangJM2014}
{Wang}, J.-M., {Du}, P., {Hu}, C., {et~al.} 2014, \apj, 793, 108, \dodoi{10.1088/0004-637X/793/2/108}

\bibitem[{{Wang} {et~al.}(2024){Wang}, {Lin}, {Zhang}, \& {Zhu}}]{WangYH2024}
{Wang}, Y., {Lin}, D. N.~C., {Zhang}, B., \& {Zhu}, Z. 2024, \apjl, 962, L7, \dodoi{10.3847/2041-8213/ad20e5}

\bibitem[{{Wevers} {et~al.}(2023){Wevers}, {Coughlin}, {Pasham}, {Guolo}, {Sun}, {Wen}, {Jonker}, {Zabludoff}, {Malyali}, {Arcodia}, {Liu}, {Merloni}, {Rau}, {Grotova}, {Short}, \& {Cao}}]{Wevers2023}
{Wevers}, T., {Coughlin}, E.~R., {Pasham}, D.~R., {et~al.} 2023, \apjl, 942, L33, \dodoi{10.3847/2041-8213/ac9f36}

\bibitem[{{Whalen} {et~al.}(2006){Whalen}, {Laurent-Muehleisen}, {Moran}, \& {Becker}}]{Whalen2006}
{Whalen}, D.~J., {Laurent-Muehleisen}, S.~A., {Moran}, E.~C., \& {Becker}, R.~H. 2006, \aj, 131, 1948, \dodoi{10.1086/500825}

\bibitem[{{White} {et~al.}(2000){White}, {Becker}, {Gregg}, {Laurent-Muehleisen}, {Brotherton}, {Impey}, {Petry}, {Foltz}, {Chaffee}, {Richards}, {Oegerle}, {Helfand}, {McMahon}, \& {Cabanela}}]{White2000}
{White}, R.~L., {Becker}, R.~H., {Gregg}, M.~D., {et~al.} 2000, \apjs, 126, 133, \dodoi{10.1086/313300}

\bibitem[{{Wiseman} {et~al.}(2024){Wiseman}, {Williams}, {Arcavi}, {Galbany}, {Graham}, {H{\"o}nig}, {Newsome}, {Subrayan}, {Sullivan}, {Wang}, {Ili{\'c}}, {Nicholl}, {Oates}, {Petrushevska}, \& {Smith}}]{WisemanP2024}
{Wiseman}, P., {Williams}, R.~D., {Arcavi}, I., {et~al.} 2024, arXiv e-prints, arXiv:2406.11552, \dodoi{10.48550/arXiv.2406.11552}

\bibitem[{Wolfe(1978)}]{Wolfe1978}
Wolfe, A.~M. 1978, Pittsburgh Conference on BL LAC Objects, Held at the University of Pittsburgh, April 24-26, 1978: [Papers] No. xiii, 428 p. (Pittsburgh: {University of Pittsburgh, Dept. of Physics and Astronomy})

\bibitem[{{Woo} {et~al.}(2010){Woo}, {Treu}, {Barth}, {Wright}, {Walsh}, {Bentz}, {Martini}, {Bennert}, {Canalizo}, {Filippenko}, {Gates}, {Greene}, {Li}, {Malkan}, {Stern}, \& {Minezaki}}]{Woo2010}
{Woo}, J.-H., {Treu}, T., {Barth}, A.~J., {et~al.} 2010, \apj, 716, 269, \dodoi{10.1088/0004-637X/716/1/269}

\bibitem[{{Wu} \& {Yuan}(2018)}]{Wu2018}
{Wu}, X.-J., \& {Yuan}, Y.-F. 2018, \mnras, 479, 1569, \dodoi{10.1093/mnras/sty1423}

\bibitem[{{Yang} {et~al.}(2019){Yang}, {Shen}, {Liu}, {Wu}, { }, {Shangguan}, {Graham}, \& {Yao}}]{Yang2019}
{Yang}, Q., {Shen}, Y., {Liu}, X., {et~al.} 2019, \apj, 885, 110, \dodoi{10.3847/1538-4357/ab481a}

\bibitem[{{Yao} {et~al.}(2023){Yao}, {Ravi}, {Gezari}, {van Velzen}, {Lu}, {Schulze}, {Somalwar}, {Kulkarni}, {Hammerstein}, {Nicholl}, {Graham}, {Perley}, {Cenko}, {Stein}, {Ricarte}, {Chadayammuri}, {Quataert}, {Bellm}, {Bloom}, {Dekany}, {Drake}, {Groom}, {Mahabal}, {Prince}, {Riddle}, {Rusholme}, {Sharma}, {Sollerman}, \& {Yan}}]{Yao2023}
{Yao}, Y., {Ravi}, V., {Gezari}, S., {et~al.} 2023, \apjl, 955, L6, \dodoi{10.3847/2041-8213/acf216}

\bibitem[{{Zacharias} {et~al.}(2017){Zacharias}, {B{\"o}ttcher}, {Jankowsky}, {Lenain}, {Wagner}, \& {Wierzcholska}}]{Zacharias2017}
{Zacharias}, M., {B{\"o}ttcher}, M., {Jankowsky}, F., {et~al.} 2017, \apj, 851, 72, \dodoi{10.3847/1538-4357/aa9bee}

\bibitem[{{Zhong} {et~al.}(2022){Zhong}, {Li}, {Berczik}, \& {Spurzem}}]{zhongsy2022}
{Zhong}, S., {Li}, S., {Berczik}, P., \& {Spurzem}, R. 2022, \apj, 933, 96, \dodoi{10.3847/1538-4357/ac71ad}

\bibitem[{{Zhou} {et~al.}(2006){Zhou}, {Wang}, {Yuan}, {Lu}, {Dong}, {Wang}, \& {Lu}}]{zhouhongyan2006}
{Zhou}, H., {Wang}, T., {Yuan}, W., {et~al.} 2006, \apjs, 166, 128, \dodoi{10.1086/504869}

\bibitem[{{Zhuang} \& {Ho}(2020)}]{ZhuangMY2020}
{Zhuang}, M.-Y., \& {Ho}, L.~C. 2020, \apj, 896, 108, \dodoi{10.3847/1538-4357/ab8f2e}

\bibitem[{{Zu} {et~al.}(2013){Zu}, {Kochanek}, {Koz{\l}owski}, \& {Udalski}}]{ZUying2013}
{Zu}, Y., {Kochanek}, C.~S., {Koz{\l}owski}, S., \& {Udalski}, A. 2013, \apj, 765, 106, \dodoi{10.1088/0004-637X/765/2/106}

\end{thebibliography}

\end{document}